\documentclass[twocolumn:]{pasj02}
\usepackage[switch,mathlines]{lineno} %linenumber    

\Received{}%{yyyy/mm/dd}
\Accepted{}%{yyyy/mm/dd}
\usepackage{amsmath}

\usepackage[whole]{bxcjkjatype}

\begin{document} 

\title{ 
%\LETTERLABEL %%% <-- uncomment for LETTER article  
%\REVIEWLABEL %%% <-- uncomment for REVIEW article  
XRISM/Resolve View of Abell 2319: Turbulence, Sloshing, and ICM Dynamics}

%%% begin:list of authors
% Do NOT capitalize all letters in "textsc".

\author{
XRISM \textsc{Collaboration},
Marc \textsc{Audard},\altaffilmark{1} 
Hisamitsu \textsc{Awaki},\altaffilmark{2} 
Ralf \textsc{Ballhausen},\altaffilmark{3,4,5} \orcid{0000-0002-1118-8470}
Aya \textsc{Bamba},\altaffilmark{6} \orcid{0000-0003-0890-4920}
Ehud \textsc{Behar},\altaffilmark{7} \orcid{0000-0001-9735-4873}
Rozenn \textsc{Boissay-Malaquin},\altaffilmark{8,4,5} \orcid{0000-0003-2704-599X}
Laura \textsc{Brenneman},\altaffilmark{9} \orcid{0000-0003-2663-1954}
Gregory V. \textsc{Brown},\altaffilmark{10} \orcid{0000-0001-6338-9445}
Lia \textsc{Corrales},\altaffilmark{11} \orcid{0000-0002-5466-3817}
Elisa \textsc{Costantini},\altaffilmark{12} \orcid{0000-0001-8470-749X}
Renata \textsc{Cumbee},\altaffilmark{4} \orcid{0000-0001-9894-295X}
Maria \textsc{Diaz Trigo},\altaffilmark{13} \orcid{0000-0001-7796-4279}
Chris \textsc{Done},\altaffilmark{14} \orcid{0000-0002-1065-7239}
Tadayasu \textsc{Dotani},\altaffilmark{15} 
Ken \textsc{Ebisawa},\altaffilmark{15} \orcid{0000-0002-5352-7178}
Megan E. \textsc{Eckart},\altaffilmark{10} \orcid{0000-0003-3894-5889}
Dominique \textsc{Eckert},\altaffilmark{1} \orcid{0000-0001-7917-3892}
Satoshi \textsc{Eguchi},\altaffilmark{16} \orcid{0000-0003-2814-9336}
Teruaki \textsc{Enoto},\altaffilmark{17} \orcid{0000-0003-1244-3100}
Yuichiro \textsc{Ezoe},\altaffilmark{18} 
Adam \textsc{Foster},\altaffilmark{9} \orcid{0000-0003-3462-8886}
Ryuichi \textsc{Fujimoto},\altaffilmark{15} \orcid{0000-0002-2374-7073}
Yutaka \textsc{Fujita},\altaffilmark{18} \orcid{0000-0003-0058-9719}
Yasushi \textsc{Fukazawa},\altaffilmark{19} \orcid{0000-0002-0921-8837}
Kotaro \textsc{Fukushima},\altaffilmark{15} \orcid{0000-0001-8055-7113}
Akihiro \textsc{Furuzawa},\altaffilmark{20} 
Luigi \textsc{Gallo},\altaffilmark{21} \orcid{0009-0006-4968-7108}
Javier \textsc{Garc\'ia},\altaffilmark{4,22} \orcid{0000-0003-3828-2448}
Liyi \textsc{Gu},\altaffilmark{12} \orcid{0000-0001-9911-7038}
Matteo \textsc{Guainazzi},\altaffilmark{23} \orcid{0000-0002-1094-3147}
Kouichi \textsc{Hagino},\altaffilmark{6} \orcid{0000-0003-4235-5304}
Kenji \textsc{Hamaguchi},\altaffilmark{8,4,5} \orcid{0000-0001-7515-2779}
Isamu \textsc{Hatsukade},\altaffilmark{24} \orcid{0000-0003-3518-3049}
Katsuhiro \textsc{Hayashi},\altaffilmark{15} \orcid{0000-0001-6922-6583}
Takayuki \textsc{Hayashi},\altaffilmark{8,4,5} \orcid{0000-0001-6665-2499}
Natalie \textsc{Hell},\altaffilmark{10} \orcid{0000-0003-3057-1536}
Edmund \textsc{Hodges-Kluck},\altaffilmark{4}\orcid{0000-0002-2397-206X}
Ann \textsc{Hornschemeier},\altaffilmark{4} \orcid{0000-0001-8667-2681}
Yuto \textsc{Ichinohe},\altaffilmark{25} \orcid{0000-0002-6102-1441}
Daiki \textsc{Ishi},\altaffilmark{15}
Manabu \textsc{Ishida},\altaffilmark{15} 
Kumi \textsc{Ishikawa},\altaffilmark{18} 
Yoshitaka \textsc{Ishisaki},\altaffilmark{18} 
Jelle \textsc{Kaastra},\altaffilmark{12,26} \orcid{0000-0001-5540-2822}
Timothy \textsc{Kallman},\altaffilmark{4} 
Erin \textsc{Kara},\altaffilmark{27} \orcid{0000-0003-0172-0854}
Satoru \textsc{Katsuda},\altaffilmark{28} \orcid{0000-0002-1104-7205}
Yoshiaki \textsc{Kanemaru},\altaffilmark{15} \orcid{0000-0002-4541-1044}
Richard \textsc{Kelley},\altaffilmark{4} \orcid{0009-0007-2283-3336}
Caroline \textsc{Kilbourne},\altaffilmark{4} \orcid{0000-0001-9464-4103}
Shunji \textsc{Kitamoto},\altaffilmark{29} \orcid{0000-0001-8948-7983}
Shogo B.\textsc{Kobayashi},\altaffilmark{30} \orcid{0000-0001-7773-9266}
Takayoshi \textsc{Kohmura},\altaffilmark{31} 
Aya \textsc{Kubota},\altaffilmark{32} 
Maurice \textsc{Leutenegger},\altaffilmark{4} \orcid{0000-0002-3331-7595}
Michael \textsc{Loewenstein},\altaffilmark{3,4,5} \orcid{0000-0002-1661-4029}
Yoshitomo \textsc{Maeda},\altaffilmark{15} \orcid{0000-0002-9099-5755}
Maxim \textsc{Markevitch},\altaffilmark{4} \orcid{0000-0003-0144-4052}
Hironori \textsc{Matsumoto},\altaffilmark{33} 
Kyoko \textsc{Matsushita},\altaffilmark{30} \orcid{0000-0003-2907-0902}
Dan \textsc{McCammon},\altaffilmark{34} \orcid{0000-0001-5170-4567}
Brian \textsc{McNamara},\altaffilmark{35} 
Francois \textsc{Mernier},\altaffilmark{3,4,5} \orcid{0000-0002-7031-4772}
Eric \textsc{Miller},\altaffilmark{27} \orcid{0000-0002-3031-2326}
Jon \textsc{Miller},\altaffilmark{11} \orcid{0000-0003-2869-7682}
Ikuyuki \textsc{Mitsuishi},\altaffilmark{36}\orcid{0000-0002-9901-233X}
Misaki \textsc{Mizumoto},\altaffilmark{37} \orcid{0000-0003-2161-0361}
Tsunefumi \textsc{Mizuno},\altaffilmark{38} \orcid{0000-0001-7263-0296}
Koji \textsc{Mori},\altaffilmark{24} \orcid{0000-0002-0018-0369}
Koji \textsc{Mukai},\altaffilmark{8,4,5} \orcid{0000-0002-8286-8094}
Hiroshi \textsc{Murakami},\altaffilmark{39} 
Richard \textsc{Mushotzky},\altaffilmark{3} \orcid{0000-0002-7962-5446}
Hiroshi \textsc{Nakajima},\altaffilmark{40} \orcid{0000-0001-6988-3938}
Kazuhiro \textsc{Nakazawa},\altaffilmark{36}\altemailmark  \orcid{0000-0003-2930-350X}
Jan-Uwe \textsc{Ness},\altaffilmark{41} 
Kumiko \textsc{Nobukawa},\altaffilmark{42} \orcid{0000-0002-0726-7862}
Masayoshi \textsc{Nobukawa},\altaffilmark{43} \orcid{0000-0003-1130-5363}
Hirofumi \textsc{Noda},\altaffilmark{44} \orcid{0000-0001-6020-517X}
Hirokazu \textsc{Odaka},\altaffilmark{33} 
Shoji \textsc{Ogawa},\altaffilmark{15} \orcid{0000-0002-5701-0811}
Anna \textsc{Ogorzalek},\altaffilmark{3,4,5} \orcid{0000-0003-4504-2557}
Takashi \textsc{Okajima},\altaffilmark{4} \orcid{0000-0002-6054-3432}
Naomi \textsc{Ota},\altaffilmark{45} \orcid{0000-0002-2784-3652}
Stephane \textsc{Paltani},\altaffilmark{1} \orcid{0000-0002-8108-9179}
Robert \textsc{Petre},\altaffilmark{4} \orcid{0000-0003-3850-2041}
Paul \textsc{Plucinsky},\altaffilmark{9} \orcid{0000-0003-1415-5823}
Frederick \textsc{Porter},\altaffilmark{4}\altemailmark \orcid{0000-0002-6374-1119}
Katja \textsc{Pottschmidt},\altaffilmark{8,4,5} \orcid{0000-0002-4656-6881}
Kosuke \textsc{Sato},\altaffilmark{46} \orcid{0000-0001-5774-1633}
Toshiki \textsc{Sato},\altaffilmark{47} 
Makoto \textsc{Sawada},\altaffilmark{29} \orcid{0000-0003-2008-6887}
Hiromi \textsc{Seta},\altaffilmark{18} 
Megumi \textsc{Shidatsu},\altaffilmark{2} \orcid{0000-0001-8195-6546}
Aurora \textsc{Simionescu},\altaffilmark{12} \orcid{0000-0002-9714-3862}
Randall \textsc{Smith},\altaffilmark{9} \orcid{0000-0003-4284-4167}
Hiromasa \textsc{Suzuki},\altaffilmark{24} \orcid{0000-0002-8152-6172}
Andrew \textsc{Szymkowiak},\altaffilmark{48} \orcid{0000-0002-4974-687X}
Hiromitsu \textsc{Takahashi},\altaffilmark{19} \orcid{0000-0001-6314-5897}
Mai \textsc{Takeo},\altaffilmark{49} 
Toru \textsc{Tamagawa},\altaffilmark{25} 
Keisuke \textsc{Tamura},\altaffilmark{8,4,5} 
Takaaki \textsc{Tanaka},\altaffilmark{50} \orcid{0000-0002-4383-0368}
Atsushi \textsc{Tanimoto},\altaffilmark{51} \orcid{0000-0002-0114-5581}
Makoto \textsc{Tashiro},\altaffilmark{28,15} \orcid{0000-0002-5097-1257}
Yukikatsu \textsc{Terada},\altaffilmark{28,15} \orcid{0000-0002-2359-1857}
Yuichi \textsc{Terashima},\altaffilmark{2} \orcid{0000-0003-1780-5481}
Yohko \textsc{Tsuboi},\altaffilmark{52} 
Masahiro \textsc{Tsujimoto},\altaffilmark{15} \orcid{0000-0002-9184-5556}
Hiroshi \textsc{Tsunemi},\altaffilmark{33} 
Takeshi \textsc{Tsuru},\altaffilmark{17} \orcid{0000-0002-5504-4903}
Hiroyuki \textsc{Uchida},\altaffilmark{17} \orcid{0000-0003-1518-2188}
Nagomi \textsc{Uchida},\altaffilmark{15} \orcid{0000-0002-5641-745X}
Yuusuke \textsc{Uchida},\altaffilmark{31}\altemailmark \orcid{0000-0002-7962-4136}
Hideki \textsc{Uchiyama},\altaffilmark{53} \orcid{0000-0003-4580-4021}
Yoshihiro \textsc{Ueda},\altaffilmark{54} \orcid{0000-0001-7821-6715}
Shinichiro \textsc{Uno},\altaffilmark{55} 
Jacco \textsc{Vink},\altaffilmark{56,12} \orcid{0000-0002-4708-4219}
Shin \textsc{Watanabe},\altaffilmark{15} \orcid{0000-0003-0441-7404}
Brian J.\ \textsc{Williams},\altaffilmark{4} \orcid{0000-0003-2063-381X}
Satoshi \textsc{Yamada},\altaffilmark{57} \orcid{0000-0002-9754-3081}
Shinya \textsc{Yamada},\altaffilmark{29} \orcid{0000-0003-4808-893X}
Hiroya \textsc{Yamaguchi},\altaffilmark{15} \orcid{0000-0002-5092-6085}
Kazutaka \textsc{Yamaoka},\altaffilmark{36} \orcid{0000-0003-3841-0980}
Noriko \textsc{Yamasaki},\altaffilmark{15} \orcid{0000-0003-4885-5537}
Makoto \textsc{Yamauchi},\altaffilmark{24} \orcid{0000-0003-1100-1423}
Shigeo \textsc{Yamauchi},\altaffilmark{58} 
Tahir \textsc{Yaqoob},\altaffilmark{8,4,5} 
Tomokage \textsc{Yoneyama},\altaffilmark{52} 
Tessei \textsc{Yoshida},\altaffilmark{15} 
Mihoko \textsc{Yukita},\altaffilmark{59,4} \orcid{0000-0001-6366-3459}
Irina \textsc{Zhuravleva},\altaffilmark{60} \orcid{0000-0001-7630-8085}
Riccardo~\textsc{Seppi}\altaffilmark{1},
Itsuki~\textsc{Aihara}\altaffilmark{30}, and 
Yuki~\textsc{Omiya}\altaffilmark{36}\altemailmark \orcid{0009-0009-9196-4174}
}

\altaffiltext{1}{Department of Astronomy, University of Geneva, Versoix CH-1290, Switzerland}
\altaffiltext{2}{Department of Physics, Ehime University, Ehime 790-8577, Japan}
\altaffiltext{3}{Department of Astronomy, University of Maryland, College Park, MD 20742, USA}
\altaffiltext{4}{NASA / Goddard Space Flight Center, Greenbelt, MD 20771, USA}
\altaffiltext{5}{Center for Research and Exploration in Space Science and Technology, NASA / GSFC (CRESST II), Greenbelt, MD 20771, USA}
\altaffiltext{6}{Department of Physics, University of Tokyo, Tokyo 113-0033, Japan}
\altaffiltext{7}{Department of Physics, Technion, Technion City, Haifa 3200003, Israel}
\altaffiltext{8}{Center for Space Sciences and Technology, University of Maryland, Baltimore County (UMBC), Baltimore, MD, 21250 USA}
\altaffiltext{9}{Center for Astrophysics | Harvard-Smithsonian, Cambridge, MA 02138, USA}
\altaffiltext{10}{Lawrence Livermore National Laboratory, Livermore, CA 94550, USA}
\altaffiltext{11}{Department of Astronomy, University of Michigan, Ann Arbor, MI 48109, USA}
\altaffiltext{12}{SRON Netherlands Institute for Space Research, Leiden, The Netherlands}
\altaffiltext{13}{ESO, Karl-Schwarzschild-Strasse 2, 85748, Garching bei München, Germany}
\altaffiltext{14}{Centre for Extragalactic Astronomy, Department of Physics, University of Durham, Durham DH1 3LE, UK}
\altaffiltext{15}{Institute of Space and Astronautical Science (ISAS), Japan Aerospace Exploration Agency (JAXA), Kanagawa 252-5210, Japan}
\altaffiltext{16}{Department of Economics, Kumamoto Gakuen University, Kumamoto 862-8680 Japan}
\altaffiltext{17}{Department of Physics, Kyoto University, Kyoto 603-8047, Japan}
\altaffiltext{18}{Department of Physics, Tokyo Metropolitan University, Tokyo 192-0397, Japan}
\altaffiltext{19}{Department of Physics, Hiroshima University, Hiroshima 739-8526, Japan}
\altaffiltext{20}{Department of Physics, Fujita Health University, Aichi 470-1192, Japan}
\altaffiltext{21}{Department of Astronomy and Physics, Saint Mary’s University, Nova Scotia B3H 3C3, Canada}
\altaffiltext{22}{California Institute of Technology, Pasadena, CA 91125, USA}
\altaffiltext{23}{European Space Agency (ESA), European Space Research and Technology Centre (ESTEC), 2200 AG Noordwijk, The Netherlands}
\altaffiltext{24}{Faculty of Engineering, University of Miyazaki, 1-1 Gakuen-Kibanadai-Nishi, Miyazaki, Miyazaki 889-2192, Japan}
\altaffiltext{25}{RIKEN Nishina Center, Saitama 351-0198, Japan}
\altaffiltext{26}{Leiden Observatory, University of Leiden, P.O. Box 9513, NL-2300 RA, Leiden, The Netherlands}
\altaffiltext{27}{Kavli Institute for Astrophysics and Space Research, Massachusetts Institute of Technology, MA 02139, USA}
\altaffiltext{28}{Department of Physics, Saitama University, Saitama 338-8570, Japan}
\altaffiltext{29}{Department of Physics, Rikkyo University, Tokyo 171-8501, Japan}
\altaffiltext{30}{Faculty of Physics, Tokyo University of Science, Tokyo 162-8601, Japan}
\altaffiltext{31}{Faculty of Science and Technology, Tokyo University of Science, Chiba 278-8510, Japan}
\altaffiltext{32}{Department of Electronic Information Systems, Shibaura Institute of Technology, Saitama 337-8570, Japan}
\altaffiltext{33}{Department of Earth and Space Science, Osaka University, Osaka 560-0043, Japan}
\altaffiltext{34}{Department of Physics, University of Wisconsin, WI 53706, USA}
\altaffiltext{35}{Department of Physics \& Astronomy, Waterloo Centre for Astrophysics, University of Waterloo, Ontario N2L 3G1, Canada}
\altaffiltext{36}{Graduate School of Science, Nagoya University, Aichi 464-8602, Japan}
\altaffiltext{37}{Science Research Education Unit, University of Teacher Education Fukuoka, Fukuoka 811-4192, Japan}
\altaffiltext{38}{Hiroshima Astrophysical Science Center, Hiroshima University, Hiroshima 739-8526, Japan}
\altaffiltext{39}{Department of Data Science, Tohoku Gakuin University, Miyagi 984-8588}
\altaffiltext{40}{College of Science and Engineering, Kanto Gakuin University, Kanagawa 236-8501, Japan}
\altaffiltext{41}{European Space Agency (ESA), European Space Astronomy Centre (ESAC), E-28692 Madrid, Spain}
\altaffiltext{42}{Department of Science, Faculty of Science and Engineering, KINDAI University, Osaka 577-8502, JAPAN}
\altaffiltext{43}{Department of Teacher Training and School Education, Nara University of Education, Nara 630-8528, Japan}
\altaffiltext{44}{Astronomical Institute, Tohoku University, Miyagi 980-8578, Japan}
\altaffiltext{45}{Department of Physics, Nara Women’s University, Nara 630-8506, Japan}
\altaffiltext{46}{Faculty of Science, Kyoto Sangyo University, Kyoto 603-8555, Japan}
\altaffiltext{47}{School of Science and Technology, Meiji University, Kanagawa, 214-8571, Japan}
\altaffiltext{48}{Yale Center for Astronomy and Astrophysics, Yale University, CT 06520-8121, USA}
\altaffiltext{49}{Faculty of Science, University of Toyama, Toyama 930-8555, Japan}
\altaffiltext{50}{Department of Physics, Konan University, Hyogo 658-8501, Japan}
\altaffiltext{51}{Graduate School of Science and Engineering, Kagoshima University, Kagoshima, 890-8580, Japan}
\altaffiltext{52}{Department of Physics, Chuo University, Tokyo 112-8551, Japan}
\altaffiltext{53}{Faculty of Education, Shizuoka University, Shizuoka 422-8529, Japan}
\altaffiltext{54}{Department of Astronomy, Kyoto University, Kyoto 606-8502, Japan}
\altaffiltext{55}{Nihon Fukushi University, Shizuoka 422-8529, Japan}
\altaffiltext{56}{Anton Pannekoek Institute, the University of Amsterdam, Postbus 94249, 1090 GE Amsterdam, The Netherlands}
\altaffiltext{57}{RIKEN Cluster for Pioneering Research, Saitama 351-0198, Japan}
\altaffiltext{58}{Department of Physics, Faculty of Science, Nara Women’s University, Nara 630-8506, Japan}
\altaffiltext{59}{Johns Hopkins University, MD 21218, USA}
\altaffiltext{60}{Department of Astronomy and Astrophysics, University of Chicago, Chicago, IL 60637, USA}

\email{omiya$_{-}$y@u.phys.nagoya-u.ac.jp, nakazawa@u.phys.nagoya-u.ac.jp, yuuchida@rs.tus.ac.jp, frederick.s.porter@nasa.gov}

%\author{A-Firstname \textsc{A-Familyname}\altaffilmark{1}%
%\thanks{Example: Present Address is xxxxxxxxxx}}
%\altaffiltext{1}{A-Address of Institute}
%\email{aaaaa@xxx.xxx.xx.xx}

%\author{B-Firstname \textsc{B-Familyname},\altaffilmark{2}}
%\altaffiltext{2}{B-Address of Institute}
%\email{bbbbb@xxx.xxx.xx.xx}

%%% end:list of authors

%% `\KeyWords{}' always has to be placed before ``\maketitle'' 
%%  List of Key Words:  https://academic.oup.com/pasj/pages/Pasj_Keywords 

\maketitle

\begin{abstract}
We present results from XRISM/Resolve observations of the core of the
\color{black}
galaxy cluster Abell 2319, focusing on its kinematic properties. The intracluster medium (ICM) exhibits temperatures of approximately 8~keV across the core, with a prominent cold front and a high-temperature region ($\sim$11 keV) in the northwest. 
%Spatially resolved spectroscopy revealed that while 
The average gas velocity in the 3 arcmin $\times$ 4 arcmin region around the brightest cluster galaxy (BCG) covered by two Resolve pointings is consistent with that of the BCG to within 40~km~s$^{-1}$ and we found modest average velocity dispersion of 230-250~km~s$^{-1}$. On the other hand, spatially-resolved spectroscopy reveals interesting variations.
\color{black}
A blueshift of 
up to $\sim$230~km~s$^{-1}$ is observed around the east edge of the cold front, 
where the gas with the lowest specific entropy is found.
The region further south inside the cold front shows only a small velocity difference from the BCG; however, its 
\color{black}
velocity dispersion is enhanced to $\sim$400~km~s$^{-1}$, implying the development of turbulence. 
These characteristics indicate that we are observing sloshing motion with some inclination angle following BCG
and that gas phases with different specific entropy participate in sloshing with their own velocities, as expected from simulations.
\color{black}
No significant evidence for a high-redshift ICM component associated with the subcluster Abell 2319B was found
in the region covered by the current Resolve pointings.
\color{black}
These results highlight the importance of sloshing and turbulence in shaping the internal structure of Abell 2319. Further deep observations are necessary to better understand the mixing and turbulent processes within the cluster.
\end{abstract}
%\pagewiselinenumbers % linenumber

\section{Introduction}

Massive galaxy clusters in the nearby universe, with intracluster medium (ICM) temperatures reaching approximately 10~keV, represent some of the largest gravitationally bound structures formed through the hierarchical growth of cosmic large-scale structure (e.g., \cite{2005RvMP...77..207V,2012ARA&A..50..353K}). These clusters often reside at the intersections of cosmic filaments, where frequent mergers occur, indicating ongoing mass assembly (e.g., \cite{2001ApJ...561..621R,2011ApJ...728...27O}). 
Such mergers release tremendous gravitational energy, which is dissipated into the ICM via shocks and turbulence, resulting in substantial heating (e.g., \cite{2007PhR...443....1M,2020MNRAS.495..784S,2008PASJ...60..345O}). These processes also amplify magnetic fields and reaccelerate relativistic electrons, giving rise to diffuse synchrotron radio emission such as radio halos and relics (e.g., \cite{2012A&ARv..20...54F,2014IJMPD..2330007B,2019SSRv..215...16V}). Understanding these mechanisms is crucial for quantifying the non-thermal energy content of clusters and the interplay between thermal and non-thermal components during structure formation.
Although the spatial distribution and velocity dispersion of member galaxies provide constraints on merger geometries, high-resolution X-ray observations remain essential for probing the ICM and revealing features such as shock fronts, cold fronts, sloshing structures, and bulk motions (e.g., \cite{2007PhR...443....1M,2018MNRAS.476.5591B}). These observations offer key insights into the thermodynamical properties and the dynamical states of merging clusters (e.g., \cite{2020MNRAS.497.5485Y,2022MNRAS.509.1201C,2023PASJ...75...37O}).

\begin{figure}[htb]
 \begin{center}
  \includegraphics[width=8cm]{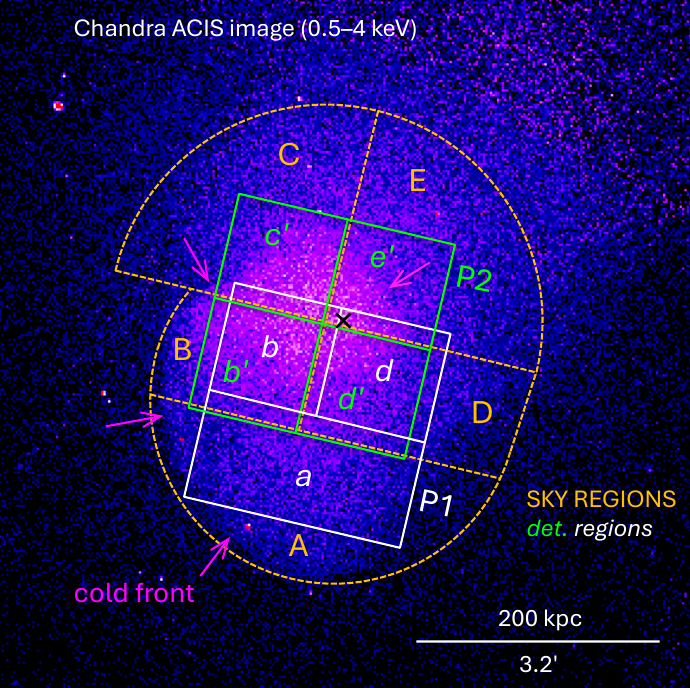} 
 \end{center}
\caption{
Chandra image of A2319 in the 0.5–4.0~keV energy band (same as used in \cite{Wang2019})
\color{black}
, overlaying regions used here for XRISM spatially-resolved spectroscopy. The boxes indicate the Resolve fields of view corresponding to observations P1 (white) and P2 (green), respectively. White boxes indicate the detector regions used for spectral extraction. For the P2 pointing, region names are labeled with a prime symbol (e.g., b′, d′) to distinguish them from those in P1. Note that Det-b and d in P1 are offset by approximately 0.5 pixel relative to Det-b′ and d′ in P2. The dashed orange regions represent the sky regions for which the ICM parameters are derived. A black ``x'' marks the position of the BCG, and purple arrows indicate the system of prominent cold fronts seen in the Chandra image.
{Alt text: Image of Abell 2319, with right ascension on the horizontal axis and declination on the vertical axis.}}\label{fig1:Chandra image}
\end{figure}

%The fluctuations in the gravitational potential arising from galaxy cluster collisions, the turbulent motion of the gas, and the incomplete mixing of gas components with different entropies are key factors that cause disturbances in the physical properties of the intracluster medium (ICM) across various spatial scales. These disturbances often manifest as fluctuations in surface brightness and shape distortions in X-ray images, contributing to the understanding of collision structures (Omiya et al. 2023). Furthermore, while direct measurement of line widths is challenging due to the energy resolution constraints of X-ray CCD cameras, careful calibration of the energy scale has enabled the measurement of bulk motion velocities in large-scale ICMs. This has facilitated ongoing discussions regarding the collision structures (Omiya et al. 2024). The XRISM satellite, launched in 2023, is capable of measuring the line-of-sight velocity of the ICM with high precision (50 km/s), providing a promising tool for offering decisive clues in understanding the geometry of collisions. Additionally, XRISM will allow for the direct observation of the motion and turbulent development within the ICM caused by collisions. The ICM is a crucial site for non-thermal energy production, and the results from XRISM are expected to be of paramount importance in advancing our understanding of these processes.

Gravitational potential fluctuations, gas turbulence, and incomplete mixing of different entropy components during a cluster merger cause disturbances in the ICM across various scales (e.g., \cite{2011ApJ...726...17P,2019MNRAS.484.4881M}). These appear as surface brightness fluctuations and distortions in X-ray images, offering insights into merging structures (e.g., \cite{2012MNRAS.421.1123C,2018ApJ...865...53Z,2024MNRAS.528.7274H,2025MNRAS.537.2198L}). 
Although direct measurement of gas motion is limited by the resolution of X-ray CCDs, careful re-calibration of energy scales have enabled 
estimates of bulk ICM velocities to $\sim 200$~km~s$^{-1}$ level \citep{2020A&A...633A..42S,2024A&A...689A.173O}.
The XRISM satellite \citep{2020SPIE11444E..22T}, launched in 2023, can measure line-of-sight ICM velocities with 
$\sim 15$~km~s$^{-1}$ accuracy,
\color{black}
offering a powerful tool for revealing merger geometry and internal ICM motion. These observations are crucial for understanding non-thermal energy production in the ICM (e.g., \cite{2025ApJ...982L...5X}).

Abell~2319 (hereafter A2319) is one of the 
%most X-ray luminous 
X-ray brightest 
nearby galaxy clusters ($z$ = 0.055, \cite{2023ApJS..267...41K}), characterized by a hot ICM with temperatures reaching $\sim$8 keV and a prominent cold front near its center (e.g., \cite{1996ApJ...465L...1M,1999ApJ...525L..73M,2004ApJ...605..695G,2024ApJ...973...98U}). 
Sunyaev-Zel'dovich effect measurements provide further evidence for the presence of hot gas in the outskirts, complementing the X-ray observations \citep{2018A&A...614A...7G,2019A&A...622A.136H}.
Optical observations reveal that the
brightest cluster \color{black} 
galaxy (BCG), CGCG 230-007, has a redshift of $z_{\text{BCG}} = 0.05458$, which matches that of the %overall 
average
redshift of the cluster \citep{1995AJ....110...32O}. These observations also suggest that the system consists of two subclusters 
superimposed along the line of sight
\color{black}
, A2319A and A2319B, indicating an ongoing merger.
\color{black}
A giant radio halo extending 
to 2~Mpc
\color{black}
further supports this scenario, although the detailed merger geometry remains under debate \citep{1997NewA....2..501F,2001A&A...369..441G,2004ApJ...605..695G,2009MNRAS.399.1307M,2013ApJ...779..189F}. 

X-ray observations from Chandra reveal enhanced surface brightness extending counterclockwise from the BCG, suggestive of sloshing motions in the plane of the sky \citep{2016MNRAS.461..684W,2021MNRAS.504.2800I}. The radio halo shows a slight southwestward extension, possibly tracing turbulence induced by large-scale gas motions from southwest to northeast \citep{2015MNRAS.448.2495S}. In contrast, the redshift of the subcluster A2319B is offset by %optical substructure analysis identified a subcluster with a line-of-sight velocity offset 
%exceeding 
 $\sim 3000$~km~s$^{-1}$ from the main cluster A2319A  \citep{1995AJ....110...32O}, implying a merger along the line of sight. These seemingly contradictory features raise the question of whether they originate from a single merger event or multiple distinct interactions. 

A2319 was within 
\color{black}
the XRISM sky visibility window right after the launch. 
Consequently, the cluster was selected as one of the first-light observation targets \color{black}
of the mission \citep{2025arXiv250320180U}. In this study, we present results from XRISM/Resolve observations of the core of A2319, focusing on its thermodynamical properties and dynamical state.
%\color{red}A2319が天球面上の衝突がどうか。\color{black}
%A2319の衝突描像はいまだに議論が行われている。A2319のChandraのX線画像は、BCG近くから反時計周りにエンハンスしており、Sloshingの運動は天球面上に近い面で起こっているように見える(Ichinohe+2020)。また、銀河団全体を覆う巨大なハローは若干南西に伸びており、粒子を加速させる乱流は南西から先に励起された、つまりこの運動は南西から北東を通って回転していると解釈されている(Storm+2015)。一方で、可視光のグルーピング解析(Oegerle+1995)は1000km/s以上のの速度差を持つサブクランプの存在を発見している。
%もし、このクランプがコールドフロントを形成しているのだとするとSloshing運動は視線上で回転していると考えられる。
%\color{magenta}
%＊：このことから、視線方向に衝突しつつある銀河団であると考えられるが、sloshingの形状は上記のように天球面上の動きを示唆しており、前者と後者が同じ２体衝突に由来するのか、別々の衝突によるものかは明らかではない。
%A2319は高温銀河団の中でICMの中心集中が強く、XRISMの観測立ち上げ時に観測可能な天空領域にあったことから、initial operation の観測ターゲットとなった。本論文ではXRISM による A2319 銀河団の観測データ解析の内容を報告する。
Throughout this paper, we use $H_0=70$ km~s$^{-1}$~Mpc$^{-1}$, $\Omega_m=0.3$ flat cosmology, in which $1'=62.3$ kpc at the cluster redshift.
All errors are in 1$\sigma$ confidence interval, otherwise noted.
\color{black}

\section{Observations and Data Reduction}

%\subsection{XRISM/Resolve}
\subsection{XRISM Observations}
%XRISM/{\it Resolve}は2024年10月13日から19日にかけてA2319の中心を3度観測しました。そのうちの2つ(obsid:000102000・000103000)は中心近く(RA,Dec=290.298,43.945)をそれぞれ140~ksと189~ksのexposure時間の間で指しており、一つ(000101000)は、コールドフロントの近くを147~ksのexposure時間の間で指しました。これらの観測は、XRISMの初期機能確認運用の間に実施されたため、長い時間でFe55 filter wheel configurationが実施されており、open filter wheel configurationでのexposure時間はそれぞれxx~ks,xx~ks,xx~ksです。
XRISM/Resolve conducted three observations of the central region A2319 between October 13 and 24, 2024. As shown in figure~\ref{fig1:Chandra image}, two of these observations (OBSID: 000102000 and 000103000) were pointed near the cluster core (P2; RA, Dec = 290.298, 43.945) for 140 ks and 189 ks, respectively. The remaining observation (OBSID: 000101000) targeted a region near the cold front (P1; RA, Dec = 290.300, 43.924) for 147 ks. 
%As these 
The three observations were carried out during the
commissioning phase of the system. Because the gate valve %was still closed, 
opening operation did not go well, 
the original gain calibration plan cannot be used and the mission needed to establish the new gain calibration procedure. As such, \color{black} 
%initial performance verification phase of XRISM, 
a significant portion of the exposure time was conducted under the $^{55}$Fe filter wheel configuration.
As the Mn-K line emission and its tail were too strong, these data cannot be used for our study. \color{black} The net exposure times with the open filter wheel configuration were approximately 49~ks, 89~ks, and 56~ks \color{black} for OBSID 000102000, 000103000, and 000101000, respectively. 
%XRISM/Xtend was not operational at the start of these observations, and it was operational only in the third one (OBSID:000103000). \color{black}

\subsection{Data reduction and screening}

%我々はFe55フィルターホイールデータで周期的ではないゲインの変動が見られているpixel27と較正用のpixel12を除いた計34pixelの完全配列のデータからスペクトルを取り出しました。これらのイベントファイルに対して、DERIV\_MAX-dependent RISE\_TIMEとSTATUS[4] スクリーニングを適応した。今回の解析では、high-resolution primary (“Hp” orITYPE = 0)イベントのみを使用した。

To reconstruct the energy scale for the observation, we used a combination of the $^{55}$Fe irradiation and, %filter wheel measurements and, 
the calibration pixel to interpolate or extrapolate the detector gain of each pixel to the nearest filter wheel measurement. 
The observations of A2319 presented here were %observed 
carried out very early in the XRISM commissioning phase, before the modern observing strategy was implemented. Thus the standard gain fiducial methodology was not employed, and specific treatment was applied.
%
%This process was performed in effective temperature space to preserve the non-linear characteristics of the energy scale. 
The detailed methodology is described in the Appendix~\ref{appendixA}.

\color{black}

We extracted spectra from a full array of 34 pixels, excluding pixel 27, which has exhibited aperiodic gain variations, and pixel 12, which is not within the field of view. %continuously illuminated by $^{55}$Fe for calibration purposes. 
For the event files, we applied DERIV\_MAX-dependent RISE\_TIME and STATUS[4] screening. In this analysis, we use only high-resolution primary (``Hp" or ITYPE = 0) events.

%4.responseの生成

%レスポンスファイルは、heasoftの6.34を使用して生成しました。rmfファイルの生成では、グレード分岐比を取得するためにスクリーニング後のLs eventsを除いた全てのグレードのイベントファイルを使用した。Lsイベントは、天体からの直接的なX線信号に由来するものを多く含んでいるので取り除いた。応答をより正確に再現するために、whichrmfオプションはelectron loss continuumやescape peakなどを考慮した"X"を選択した。
The response files were generated using HEASoft version 6.34. In creating the redistribution matrix file (RMF), we used cleaned event files to which the recommended additional screening had been applied and that included all grades except low-resolution secondary (Ls), to better reflect the Hp branching ratio. This exclusion was needed because of the large number of false Ls events associated with the on-board application of secondary-pulse detection to atypically shaped pulses such as clipped events.
To better reproduce the detector response, the whichrmf option was set to ``L''. 
\color{black}
The ancillary response file (ARF) was generated using xaarfgen with the Chandra image as the input X-ray surface brightness distribution.
\color{black}
A non-X-ray background (NXB) spectrum was extracted from a NXB database 
(v2)
\color{black}
of Resolve night-Earth data using rslnxbgen and weighting by the distribution of geomagnetic cutoff rigidity sampled during each observation.

\section{Data Analysis and Results}

%\subsection{XRISM data analysis and results}
\subsection{Spectral modeling} 

%Xspec12.12.0(\cite{Arnaud1996})を使い、対応する原子データベースAtomDB 3.0.9をプラズマモデル計算に用いてResolveスペクトルをフィッティングした。全てのアバンダンスはLodders+2009の原始太陽系値に対する相対値で与えられている。ICMの放射モデルを再現するために、我々はxspecで与えられている velocity-broadened collisional-equilibrium model ($bapec$)を採用した。これはプラズマの温度やアバンダンスや赤方偏移や速度広がりが自由パラメーターとして与えられる。赤方偏移は、これらの観測は同じ時期に観測されたため衛星の軌道運動による効果を考慮した補正(Geocentric correction)は行わないが、可視光の銀河速度と比較をするためにフィット結果の赤方偏移に対してheasoftのbarycenコマンドを用いて計算された12.42~km~s$^{-1}$を差し引くことで、太陽中心での時刻に補正した(Heliocentric correction)。速度広がりにはthermal broadeningを考慮した。乗法吸収モデル($phabs$)はWillingaleら(2013)によって決定された11.2$\times$10$^{20}$~cm$^{-2}$の水素のカラム密度にセットした。
%フィッティングの適合度の決定にはc統計(Cash+XXXX)を採用した。
%We performed spectral fitting of the Resolve data 
Spectral fitting to the Resolve data
was done using \color{black} %次のセクションと冒頭が似てしまうので。中澤
XSPEC version 12.12.0 \citep{1996ASPC..101...17A}, adopting the atomic database AtomDB version 3.0.9 for plasma model calculations. All elemental abundances are given relative to the protosolar values reported by \citet{2009LanB...4B..712L}. To model the emission from the ICM, we employed the velocity-broadened, collisional ionization equilibrium model ($bapec$) \citep{2001ApJ...556L..91S} in XSPEC. This model allows the plasma temperature, elemental abundances, redshift, and velocity broadening to be treated as free parameters.
Thermal broadening is also included. \color{black}
%Since all observations were conducted within a short period, we did not apply geocentric corrections for the satellite's orbital motion. %%これおかしいよ。地球軌道速度の補正をしない理由にはなるけど、衛星起動速度の話は短期間の観測とは関係ない。
Although correction to satellite's orbital motion was not applied, \color{black}
to enable comparison with galaxy velocities measured in the optical band, we applied
an average
\color{black}
heliocentric correction to the fitted redshift by subtracting 12.42~km~s$^{-1}$, calculated using the barycen tool in HEASoft. %Thermal broadening was accounted for in the velocity dispersion. 
For the Galactic absorption, we used the multiplicative absorption model ($tbabs$), fixing the hydrogen column density to 11.2$\times$10$^{20}$~cm$^{-2}$, as determined by \citet{2013MNRAS.431..394W}. 
The NXB spectra of Resolve were modeled with a combination of power-law continuums and Gaussian lines, and incorporated into the spectral fitting as an additive background components.
\color{black}
The cosmic X-ray background was not considered, as it is less than 1\% of the celestial signal anywhere in the 3--9.5 keV energy band. 
Model fitting was evaluated using the C-statistic \citep{1979ApJ...228..939C}.

\begin{table}[h!]
\begin{center}
\tbl{Best-fit Parameters of full array in 3.0--9.5~keV enegy range. }{%
\begin{tabular}{@{}lccc@{}}
\hline\noalign{\vskip3pt}
\multicolumn{1}{c}{Parameter} & P1 & P2  \\ [2pt]
\hline\noalign{\vskip3pt}
 $kT$~(keV) & $8.88^{+0.32}_{-0.33}$ & $9.48^{+0.20}_{-0.20}$  \\
 Abundance~($Z_{\rm{\odot}}$)\footnotemark[$*$] & $0.42^{+0.03}_{-0.03}$ & $0.44^{+0.02}_{-0.02}$  \\
 Redshift ($\times$10$^{-2}$) & $5.4466^{+0.0079}_{-0.0099}$  & $5.4490^{+0.0048}_{-0.0053}$  \\
%  Velocity relative && \\
%  ~~~to BCG (km~s$^{-1}$) &$-36.5^{+29.7}_{-23.8}$ & $-29.7^{+14.3}_{-16.0}$\\
%  Velocity relative && \\
  Relative Velocity (km~s$^{-1}$)\footnotemark[$\dagger$] %&$-36.5^{+29.7}_{-23.8}$ & $-29.7^{+14.3}_{-16.0}$\\
  &$-37^{+30}_{-24}$ & $-30^{+14}_{-16}$\\
  Velocity dispersion (km~s$^{-1}$) & $249^{+24}_{-23}$ & $232^{+14}_{-14}$\\
  C-stat/$d.o.f$  & $13367/12994$ & $27972/25992$ \\
 [2pt]
\hline\noalign{\vskip3pt}
\end{tabular}
}\label{tab1:Fit parameter}
\begin{tabnote}
{\hbox to 0pt{\parbox{78mm}{\footnotesize
\par\noindent
\footnotemark[$*$]: We adopted the solar abundance table of \citet{2009LanB...4B..712L}. \\
\footnotemark[$\dagger$]: Velocity relative to BCG.\\
}\hss}} 
\end{tabnote}
\end{center}
\end{table}

\begin{figure*}[htpb]
 \begin{center}
  \includegraphics[width=13.2cm]{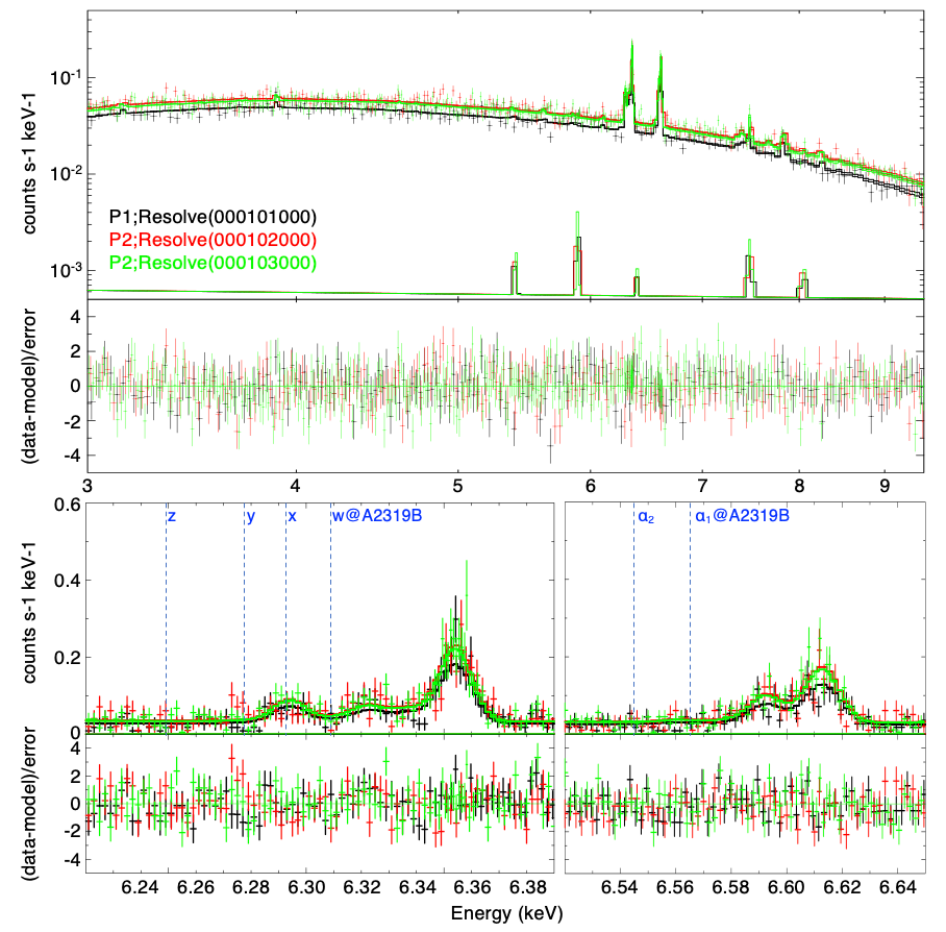} 
 \end{center}
\caption{(Top) Resolve spectra in the
3.0--9.5~keV
\color{black}
energy band for regions P1 and P2. The spectra shown in black, red, and green represent Resolve observational data corresponding to OBSIDs 000101000, 000102000, and 000103000, respectively. The best-fit models are overlaid on each spectrum 
with NXB models,
\color{black}
illustrating the goodness of fit between the observational data and the corresponding model. 
Data are binned by a factor of 4~eV for display purposes.
%Data are binned 8 counts for display with minimum significance set to 128$\sigma$. 
\color{black} 
(Bottom) Zoom-in of the Resolve spectra around the He-like Fe lines(left) and H-like Fe lines (right).
Data are binned by a factor of 2~eV for display purposes. 
\color{black}
The blue lines indicate the expected line centers of He-like Fe K$\alpha$ complex (w, x, y, z) and H-like Fe Ly$\alpha_{1,2}$ with a velocity of 18636~km~s$^{-1}$, 
which is an average velocity of A2319B.
{Alt text: Three line graphs. In the upper panel, the x axis shows the energy from 3.0 to 9.5 kilo electron volt. The y axis shows the count from 0.0005 to 0.5 counts per second and per kilo electron volt, and the residuals of minus 5 to 5 in lower part. In the lower left panel, the x axis shows the energy from 6.22 to 6.40 kilo electron volt. In the lower right panel, the x axis shows the energy from 6.50 to 6.68 kilo electron volt. The y axis shows the count from 0.0 to 0.6 counts per second and per kilo electron volt, and the residuals of minus 5 to 5 in lower part.
\color{black} 
}}\label{fig2:FoV spec}
\end{figure*}
%(TOP)領域P1およびP2における、3–7keVエネルギーバンドでのResolveおよびXtendのスペクトル。各スペクトルにはベストフィットモデルを重ねて表示しており、観測データとモデルの適合度を示している。黒と赤と緑のデータはResolveのデータを示しており、それぞれobsidが000101000・000102000・000103000のスペクトルである。青とシアンのデータはXtendのデータを示しており、それぞれ領域P1と領域P2のスペクトルである。

\subsection{Spectral fitting of full array} 
%我々は、観測P1とP2の完全配列のResolveデータを用いて、スペクトルに対してフィッティングを行いました。P2の2観測の各パラメータはリンクさせました。温度決定の統計量を向上するために、Xtendデータと同時フィッティングを行なっています。Resolveの低エネルギー側の有効面積の不定性が大きいことが報告されています。Cicely et al. 2025は、Xtendの0.5-10.0~keVのフィッティング温度とResolve温度の不一致を報告しています。そこで我々はResolve・Xtendともに3-9.5keVのエネルギー範囲でフィッティングを行いました。ベストフィットモデルを表1にリストしており、ベストフィットモデルとスペクトルを図2に示しています。

We performed spectral fitting of the full-array Resolve data from observations P1 and P2. 
Both 000102000 and 000103000 observed the same region P2,
and its difference is negligible. We therefore linked the parameters between them. 
%In addition, to improve the statistical accuracy of the temperature measurements, joint spectral fitting was conducted with the Xtend data.
%\color{red} ここではXtendではなんのパラメータをlinkしたのか明示すること。\color{black} 
%In these joint fits, the temperature and abundance parameters were linked between Resolve and Xtend to enhance the statistical constraints. On the other hand, the redshift and velocity dispersion were determined solely from the Resolve data. 
%Since Xtend lacks sufficient energy resolution and the gain calibration to constrain these parameters, 
%For simplicity, the redshift and velocity dispersion in the Xtend spectra were fixed at 0.053 and 200~km~s$^{-1}$, respectively.
\color{black}

It has been previously reported that the effective area calibration of Resolve is subject to significant uncertainties at low energy. 
%Cicely et al. (2025) pointed out a notable discrepancy between the temperatures derived from Xtend data in the 0.5–10.0 keV energy range and those obtained from Resolve. 
To mitigate this issue, we limited the fitting energy range to 3.0–-9.5 keV in this analysis.

The best-fit model parameters of the single-component fitting \color{black} are listed in table~\ref{tab1:Fit parameter}, and the spectra with the best-fit model are shown in figure~\ref{fig2:FoV spec}. Overall, the bulk ICM velocity agrees well with that of the BCG, typically within $\sim$40~km~s$^{-1}$. The ICM velocity dispersion is measured to be about 230--250~km~s$^{-1}$, which is comparable to that at the center of the Coma cluster \citep{2025ApJ...985L..20X}.
Although the velocity dispersions are similar among the two clusters, the effective length scales over which they are measured are smaller in %P1 and P2 of 
A2319 due to the presence of a denser core. Based on deprojected density and temperature profiles, we roughly estimate the effective length to be 150--200~kpc in P1 and P2 of A2319, compared to $\sim$250~kpc in the core pointing %region 
of the Coma cluster. This difference implies %stronger local pressure fluctuations and 
a more dynamically disturbed state in P1 and P2 of A2319.
\color{black}

Furthermore, a recent study using surface brightness fluctuation techniques with Chandra data reports an ICM velocity of $\sim$250~km~s$^{-1}$ within $R_{2500}$ ($\sim$700~kpc) for A2319 \citep{2024MNRAS.528.7274H}. Although the analyzed region and methodology differ from our XRISM-based approach, the derived velocity scale is in overall agreement with our results. Notably, the fluctuation-based method captures a broad range of velocity modes across spatial scales from a few tens to several hundred kiloparsecs, reinforcing the physical consistency between these independent diagnostics of the ICM velocity field.
\color{black}

%Furthermore, a recent study using surface brightness fluctuation techniques with \textit{Chandra} data reports an ICM velocity of $\sim$250~km~s$^{-1}$ within $R_{2500}$ ($\sim$700~kpc) for A2319 \citep{2024MNRAS.528.7274H}. 
%Although the region and methodology differ from our XRISM-based analysis, the derived velocity scale is in good agreement with our results, supporting the consistency between independent measurements.

%\color{magenta}
%＊：Table.1のフィット結果を見ると、20000番と30000番は統計的に優意に、「温度」「アバンダンス」が一致していない。これはちょっと面倒だね。このまま論文にするのはやはり難しい。対策が必要。\color{black}

\subsection{Spatially resolved spectroscopy} 

To quantitatively investigate the spatial structure of the ICM, we performed a spatial-spectral mixing (SSM) analysis. This method accounts for the contamination of photons from neighboring sky regions into each detector (sub-)region due to the relatively broad point spread function (PSF) of the X-ray Mirror Assembly (XMA).
We include the PSF scattering effects into our model following the multi-source ARF technique described in \citet{2018PASJ...70....9H}, using the Chandra image in the 0.5–-7.0~keV band as the input.
\color{black}

We divided the cluster core into five sky regions, labeled A through E as shown in figure~\ref{fig1:Chandra image} overlaid on the Chandra image of the cluster. The regions are selected to focus on interesting parts of the core --- for example, Reg-A isolates the ICM inside the prominent cold front. We then selected regions in the Resolve field of view that follow these sky regions as closely as possible, and extracted spectra from those regions. To account for mixing between the sky and detector regions, we used the high-resolution Chandra image as a model of the X-ray surface brightness to compute the contribution of each sky region to %each detector region. 
individual detector regions.
We performed a simultaneous fit of all spectra and model components corresponding to the individual sky regions.
To incorporate photons scattered from outside the field of view into the spectral model, the boundaries of the sky regions were extended beyond the detector field of view. Specifically, regions A, B, and D were extended to reach the cold front boundaries identified by \citet{2021MNRAS.504.2800I}, while regions C and E were defined as circular regions with radii of 3 arcmin, centered on the P2 pointing position. We note that the leak-in contributions from outside the cold front boundaries are estimated to be approximately 4\% for Reg-A, 2\% for Reg-B, and much less for other regions.  Since the parameters in these outer region cannot be reliably constrained with the current data, we simply ignored it and did not include it in the spectral modeling.

On the detector side, spectra were extracted independently for each observation. As shown in figure~\ref{fig1:Chandra image}, for the P1 pointing, we divided the field of view into regions a, b, and d, while for the P2 pointing, we used regions b′, c′, d′, and e′, resulting in seven detector regions in total. Since each detector region may receive photon contributions from all five sky regions, we generated a cross-region ARF for every combination of sky and detector regions. %To account for the small offset of approximately 0.5 pixels between the P1 and P2 pointings, ARFs were generated using the appropriate aspect solution for each ObsID. This enabled us to 
By this approach, we can 
construct a self-consistent spectral model that reflects the instrument's spatial resolution and accurately models spatial photon mixing.
Regions are selected to be wide enough in comparison with the PSF (which has a 1.3 arcmin half-power diameter) so that in each spectrum the dominant fraction of the photons originates in the underlying sky region. 
%This is illustrated in Figure \ref{fig3:spec regA}, which shows contributions into detector regions a and d+d' from the five sky regions. 

\begin{figure}[htb]
 \begin{center}
  \begin{minipage}{1.0\columnwidth}
    \centering
    \includegraphics[width=\columnwidth]{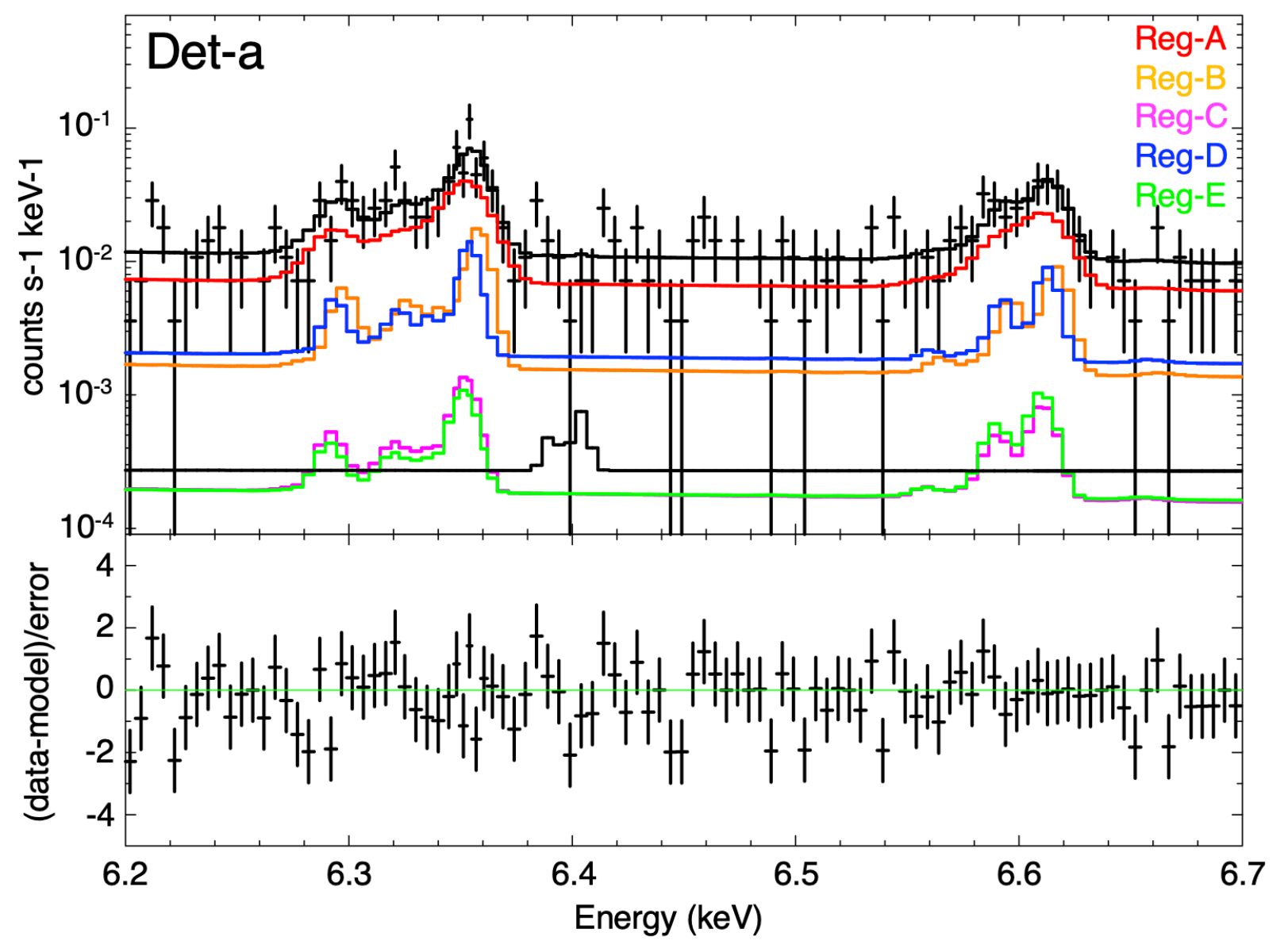}
  \end{minipage}
  %\hspace{5mm}
  \begin{minipage}{1.0\columnwidth}
    \centering
    \includegraphics[width=\columnwidth]{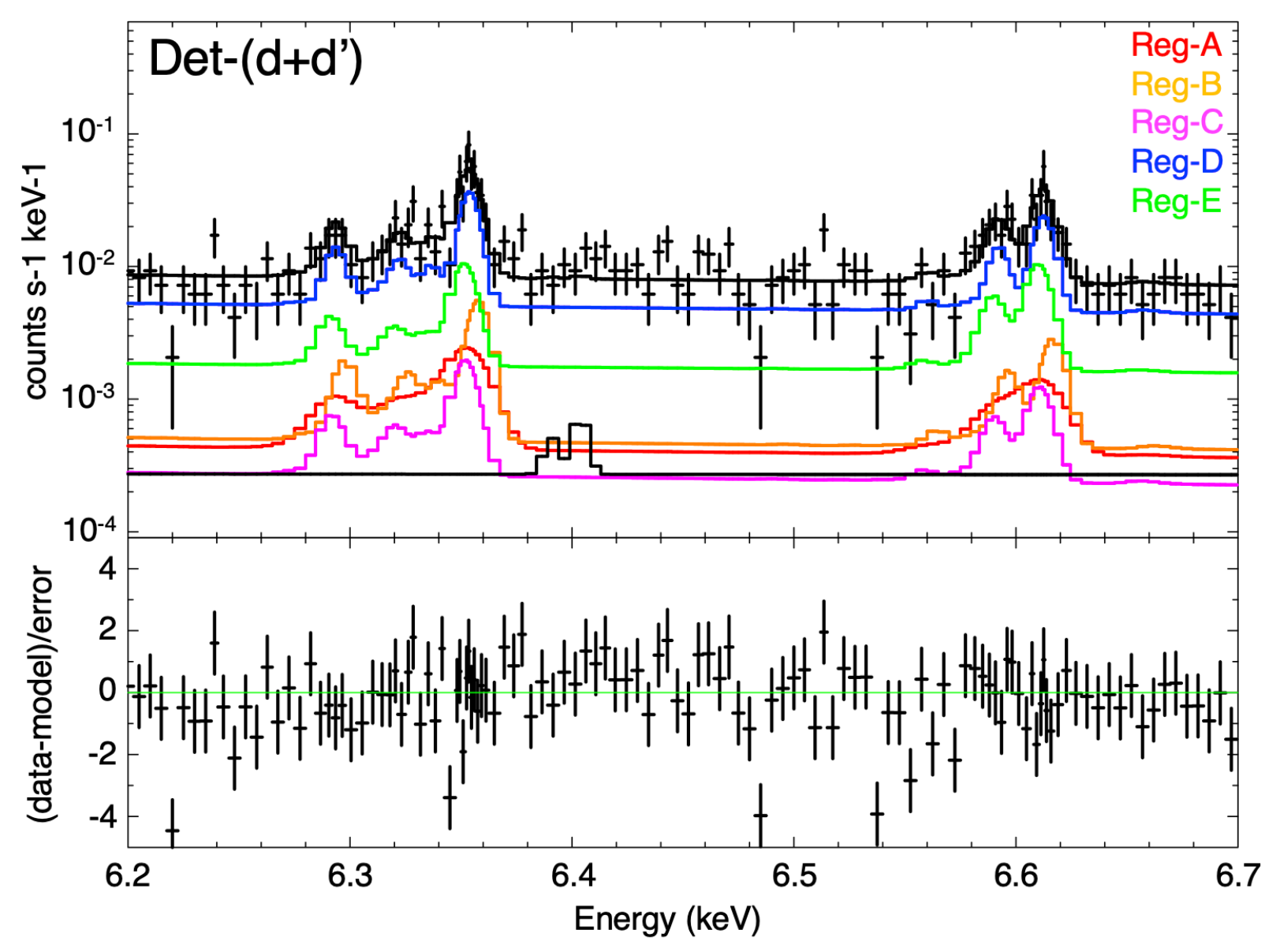}
  \end{minipage}
 \end{center}
\caption{Resolved spectra around the Fe K$\alpha$ lines of Det-a (upper) and Det-d+d' (lower) for the SSM analysis. The spectra of regions d and d′ are co-added using mathpha for display purposes. The solid lines represent the modeled contributions from individual regions: Reg-A (red), Reg-B (orange), Reg-C (magenta), Reg-D (blue), and Reg-E (green).
{Alt text: Two line graphs. In the two panels, the x axis shows the energy from 6.2 to 6.7 kilo electron volt. The y axis shows the count from 0.0009 to 0.7 counts per second and per kilo electron volt, and the residuals of minus 5 to 5 in lower part.}}
\label{fig3:spec regA}
\end{figure}

Each spectrum was simultaneously fitted using five bapec models corresponding to regions A--E, with individual components convolved with their respective ARFs. The physical parameters --   such as temperature, metal abundance, redshift, velocity dispersion and normalization -- were tied across all spectra to which the same sky region contributed.
Figure~\ref{fig3:spec regA}
\color{black}
shows the two examples of spectra in detector regions, plotted together with fitted contributions from the five sky regions.
As expected, Reg-A component is dominant in Det-a region, and Red-D component is dominant in Det-(d+d') region.

\begin{figure*}[h]
  \centering
  \begin{minipage}{1.0\columnwidth}
    \centering
    \includegraphics[width=\columnwidth]{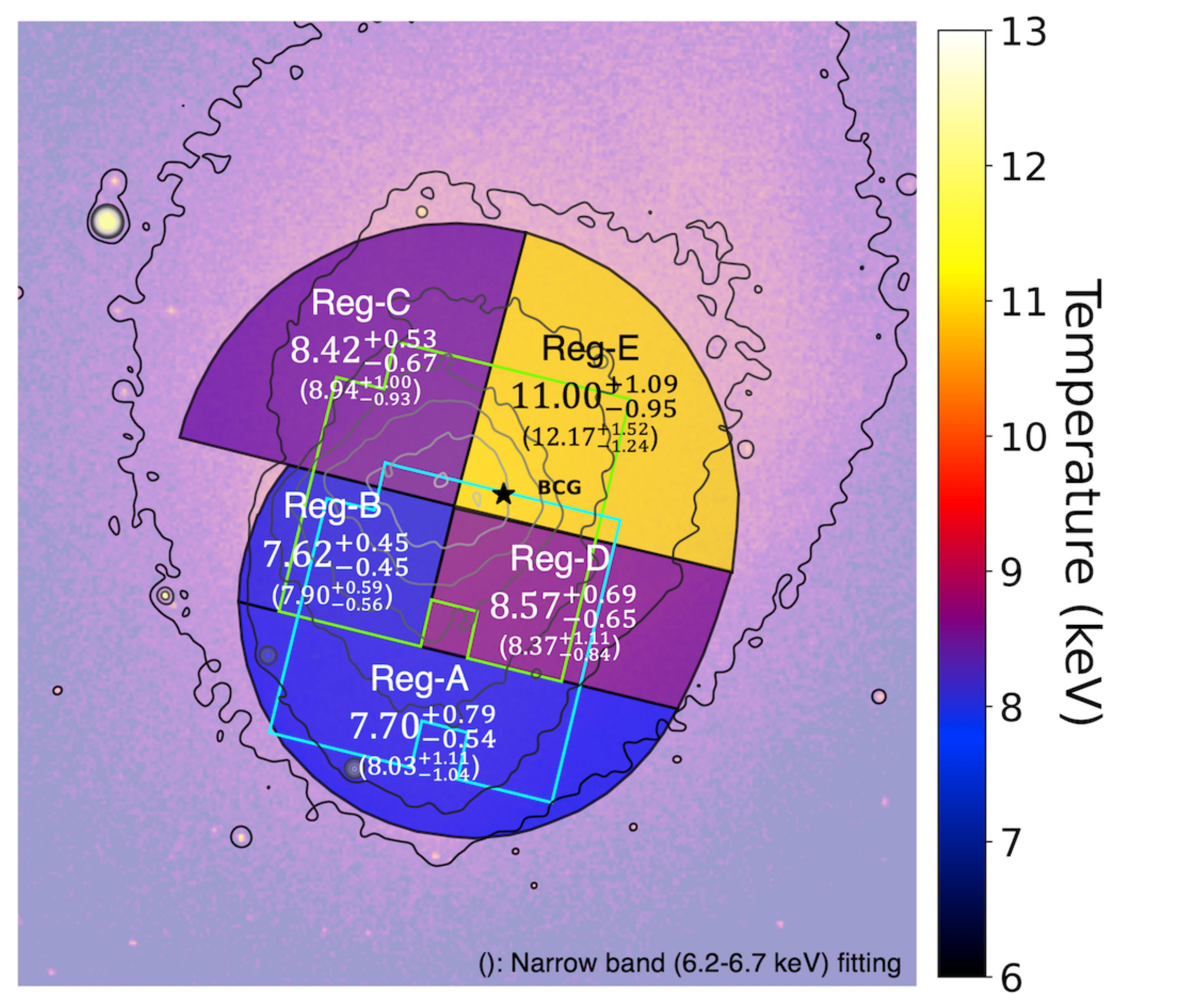}
  \end{minipage}
  %\hspace{5mm}
  \begin{minipage}{1.0\columnwidth}
    \centering
    \includegraphics[width=\columnwidth]{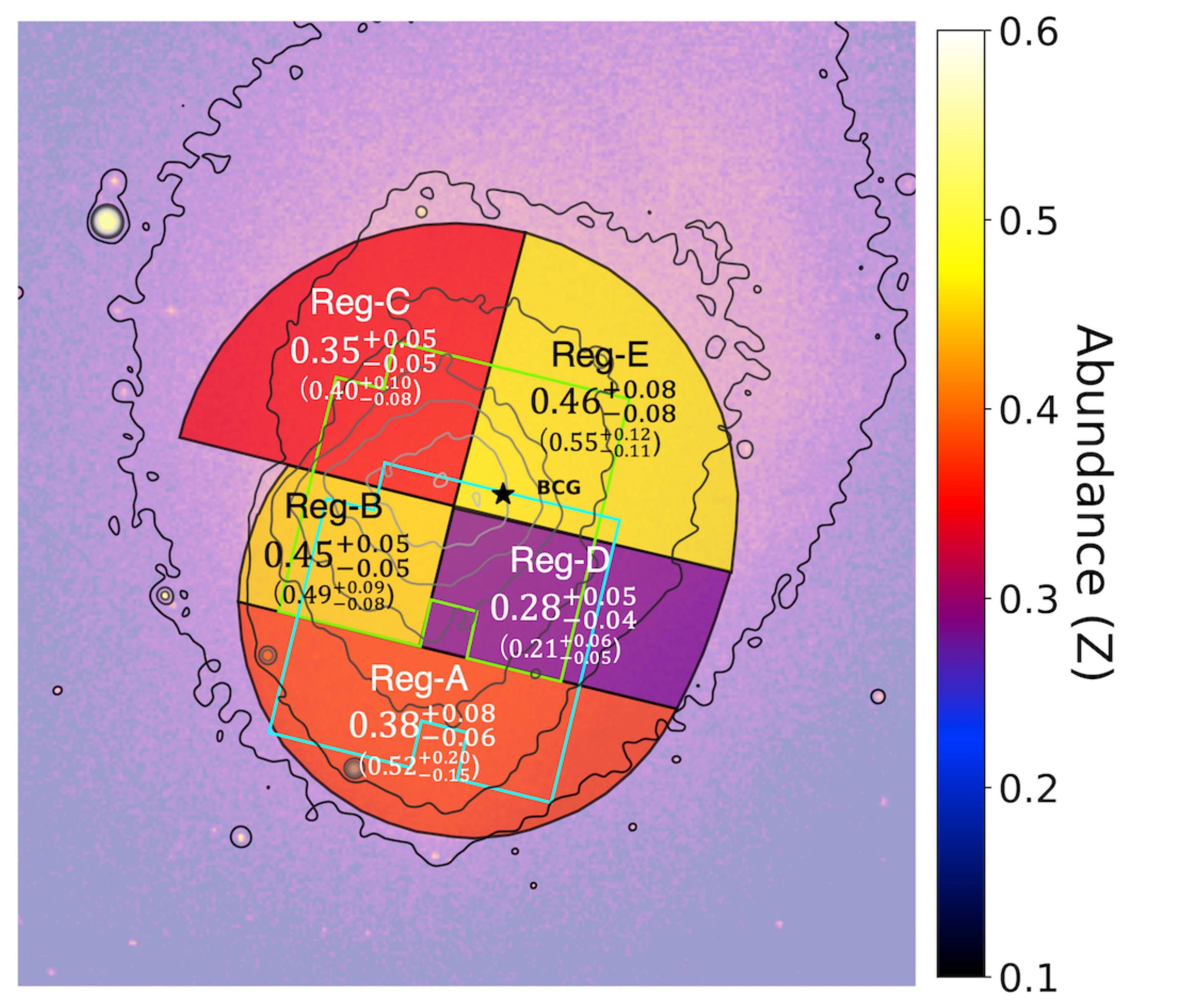}
  \end{minipage}

  %\vspace{3mm}
  
  \begin{minipage}{1.0\columnwidth}
    \centering
    \includegraphics[width=\columnwidth]{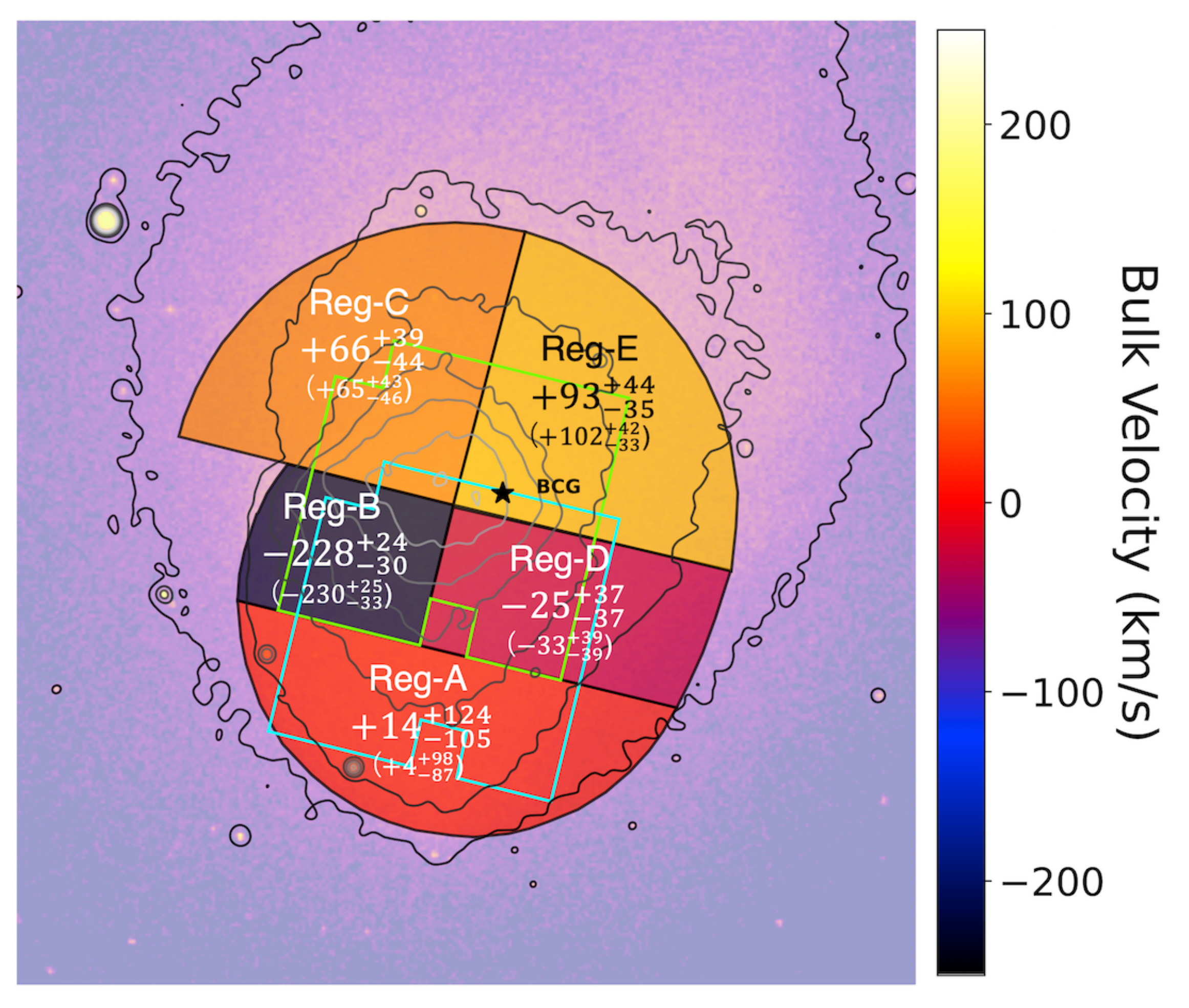}
  \end{minipage}
  %\hspace{5mm}
  \begin{minipage}{1.0\columnwidth}
    \centering
    \includegraphics[width=\columnwidth]{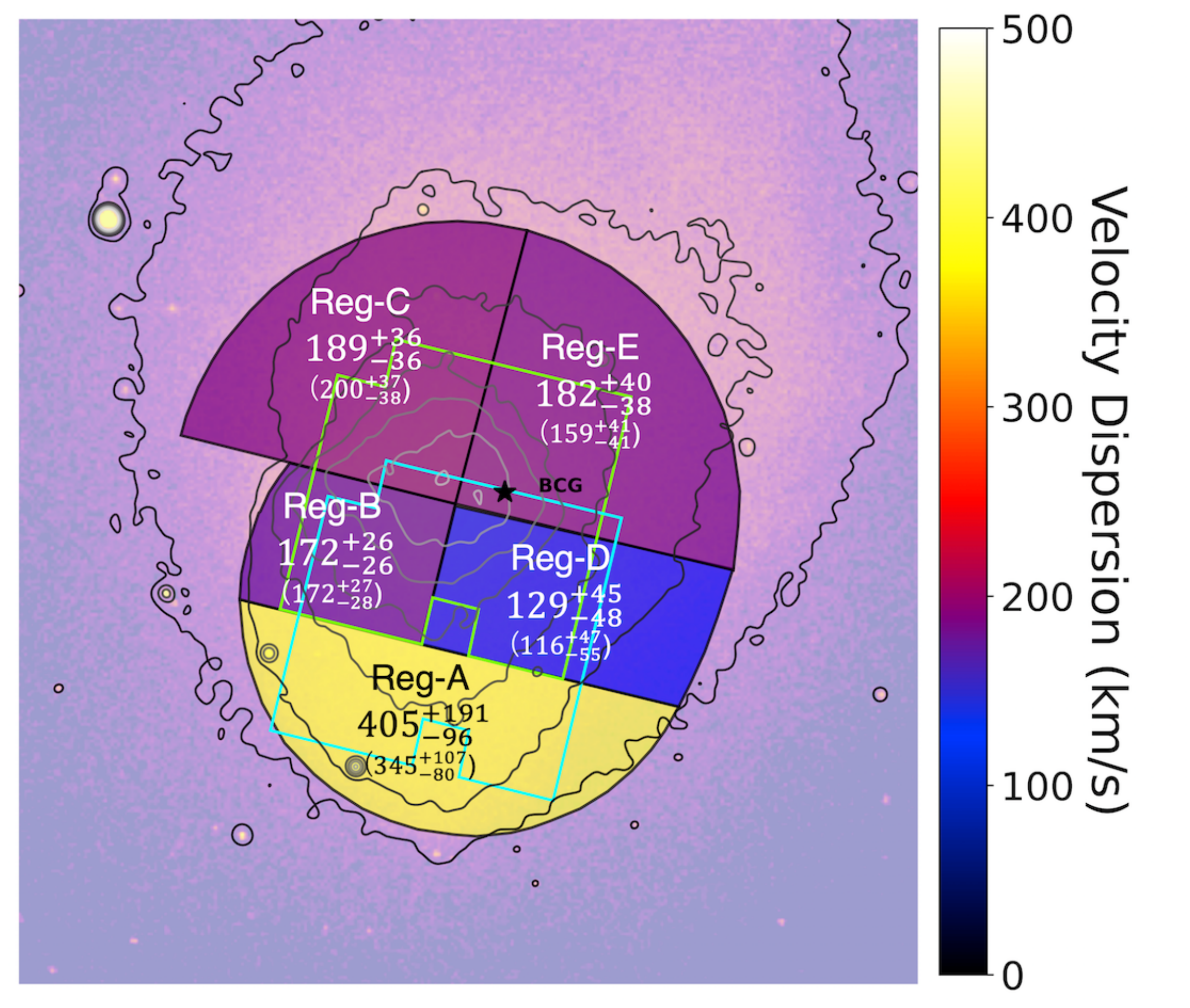}
  \end{minipage}
\caption{Maps of temperature (top left), abundance (top right), bulk velocity (bottom left), and velocity dispersion (bottom right), based on the spatial regions defined in figure~\ref{fig1:Chandra image}. 
All maps are derived from broad-band fits in the 3.0–9.5~keV range; values in parentheses indicate the corresponding results from the narrow-band fits (6.2–6.7~keV).
\color{black}
The black star indicates the position of the BCG. Contours represent the X-ray surface brightness, highlighting the correspondence with spatial structures. The boxes indicate the fields of view corresponding to observations P1 (cyan) and P2 (green). The units of the color bars are, in order:~keV, solar, km~s$^{-1}$, and km~s$^{-1}$.
{Alt text: Four color maps arranged in two rows and two columns, each with right ascension on the horizontal axis and declination on the vertical axis. The upper left map shows temperature in kilo–electronvolts ranging from 6.0 to 13.0. The upper right map shows abundance in solar units ranging from 0.1 to 0.6. The lower left map shows bulk velocity in kilometers per second ranging from minus 250 to plus 250. The lower right map shows velocity dispersion in kilometers per second ranging from 0 to 500.}}
\label{fig3:Resolve map}
\end{figure*}

The fitted results are presented in figure~\ref{fig3:Resolve map}.
For reference, the fitting results without accounting for SSM in each detector region are summarized in table~\ref{tab2:Fit parameter in detector regions} in the Appendix.
We note that, although our sky regions extend beyond the boundaries of the detector as described above, the derived ICM parameters are most relevant for the parts of the regions that are inside the field of view, where we have the data. 
If there are parameter variations within each sky region, we would not be able to detect it without additional Resolve coverage.

Regions A and D,
\color{black}
located near the cold front, exhibit bulk velocities consistent with that of the BCG, $+14^{+124}_{-105}$~km~s$^{-1}$ and 
$-25^{+37}_{-37}$~km~s$^{-1}$,
\color{black}
respectively. In contrast, Reg-B, located southeast of the BCG, exhibits a significant blueshift, $-228^{+24}_{-30}$~km~s$^{-1}$.
Meanwhile, 
\color{black}
the northern side of the BCG shows a redshift relative to the BCG: the
Reg-C exhibits a bulk velocity of 
$66^{+39}_{-44}$~km~s$^{-1}$,
\color{black}
and the Reg-E shows 
$93^{+44}_{-35}$~km~s$^{-1}$.
\color{black}
The velocity difference between Reg-B and Reg-E thus amounts to approximately 300~km~s$^{-1}$.
%The probability that the
The bulk velocities of these regions differ by
more than %are statistically consistent is $<$ 
4$\sigma$. %, suggesting a significant difference in motion.
\color{black}

%The velocity dispersion across the field generally lies in the range of $200 \pm 50$~km~s$^{–1}$, comparable to that of the Coma cluster. An enhancement is observed near the cold front: the southeastern region shows a velocity dispersion of 323$\pm$67~km~s$^{–1}$. This elevated dispersion may reflect the development of turbulence or line-of-sight mixing motions in the downstream region of the sloshing flow.
The velocity dispersion across the field generally lies in the range of %$200 \pm 50$~km~s$^{-1}$, 
100--200~km~s$^{-1}$, comparable to that of the Coma cluster. An enhancement is observed near the south end of the cold front: Reg-A shows a velocity dispersion of 
$405^{+191}_{-96}$~km~s$^{-1}$. 
\color{black}
The spectrum of Det-a around the Fe K$\alpha$ lines is shown in the upper panel of figure~\ref{fig3:spec regA}.
The difference in the line width between the high-dispersion Det-a and the low-dispersion Det-d spectra is clearly visible.
\color{black}
Notably, Reg-A is the highest velocity dispersions
reported by XRISM for the several clusters observed to date.
\color{black}
For comparison, previous studies have reported typical values of $\sim$100–200~km~s$^{-1}$ in the central regions of nearby clusters such as Perseus, Abell 2029, and Coma (e.g., \cite{2016Natur.535..117H,2025ApJ...982L...5X,2025ApJ...985L..20X}).
\color{black}

The temperature map reveals that the ICM is relatively hot across most regions. 
The central regions and the southeastern regions are slightly cooler, around 7.6--7.7 keV.
In contrast, the Reg-E exhibits higher temperatures of 
$11.0^{+1.1}_{-1.0}$~keV,
\color{black}
consistent with previous Chandra and XMM-Newton observations that identified a high-temperature belt structure \citep{2004ApJ...605..695G,2021MNRAS.504.2800I}.
The abundance map shows modest spatial variations, ranging from 
0.28 to 0.46 solar.
\color{black}
Higher abundances 
($\sim$0.45~solar)
\color{black}
are found in regions B and E, near the BCG, while somewhat lower values (0.28--0.38 solar) are observed in regions A and D.

We also performed narrow-band SSM fits restricting within 6.2-–6.7~keV range, where the temperature is primarily constrained by the Fe–K line ratio. The best-fit values from these fits are shown in parentheses in figure~\ref{fig3:spec regA}, and are consistent with those from the broadband fits. %We note that uncertainties in PSF mixing can produce apparent discontinuities in the spatial distributions of temperature and abundance; however, the derived bulk velocities and velocity dispersions remain essentially unchanged.
\color{black}

% In addition, a gradient in the abundance is observed along the cold front, decreasing in a counterclockwise direction from the BCG to the southeast.
%
%%In addition, starting from around the BCG, high abundance regions match with the counterclockwise cold front shape, to the east, and then to the south.
%
%This may reflect the redistribution of metal-enriched gas due to sloshing motions. 
\color{black}
%No strong spatial correlation is found between the abundance and velocity structures.

%Overall, the spatially resolved spectroscopy reveals that A2319’s core is dominated by sloshing motions with slightly blue- and redshifted components around the BCG, modestly enhanced turbulence near the cold front, and evidence of mixing between different entropy components, particularly in the northwestern region.

\color{black}

%% 中澤チェック 20250415ここまで

\section{Discussion}

The mean redshifts in the two XRISM/Resolve pointings covering the core of the A2319 cluster are consistent with the BCG optical redshift within 40~km~s$^{-1}$. Our spatially resolved velocity measurements uncover two significant features in this sloshing core --- (a) a $\sim$300~km~s$^{-1}$ variation in the line of sight velocities across the core, with one region (Reg-B in figure~\ref{fig3:Resolve map}) showing a highly significant $>$200~km~s$^{-1}$ deviation from the mean, while other regions showing velocities within 100~km~s$^{-1}$ of the mean, and (b) a relatively high line of sight velocity dispersion ($\sim$400~km~s$^{-1}$) in Reg-A immediately inside the cold front, while its line of sight velocity is close to the mean.
\color{black}

\subsection{%Indication of sloshing on the skyplane 
Sloshing inclined to the skyplane \color{black}}
%天球面上に水平な面でSloshingが起こっている示唆

If the low-entropy gas is rotating as a result of sloshing, the central ICM with the lowest entropy and the ICM near the cold front should be moving in three-dimensional space. 
Hydrodynamical simulations of minor mergers \citep{2016ApJ...821....6Z} between clusters demonstrated that, when the sloshing plane is oriented along the line of sight, a velocity difference amounts to $M$=0.2–0.3. Low-entropy ICM in A2319 has a temperature of $\sim$~8.5 keV and its sound speed is estimated to be $C_{s} = (\gamma k_{B}T/\mu m_{p})^{1/2} = 1500$~km~s$^{-1}$, assuming a mean molecular weight $\mu = 0.6$ and an adiabatic index $\gamma = 5/3$. In this case, a velocity difference of 300–450~km~s$^{-1}$ would be expected. %, which will be observed as %either 
%large velocity offset to the BCG 
%or
%and
%%
%large velocity dispersion. 
%\color{blue}
%Because such difference in the bulk velocity distribution is not observed, %neither of them are observed, 
%the sloshing is not mainly in the line of sight, but mainly on the sky-plane. 
%\color{black}

\color{black}
Chandra X-ray image shows a %%vortex-like structure
``spiral structure'' 
associated with sloshing motions, extending counterclockwise from the BCG to the east and then to the south, spanning approximately 200~kpc (see figure~8 of \cite{2021MNRAS.504.2800I}). 
It indicates that the sloshing motion has some motion in the skyplane.
\color{black}
%If the sloshing motion were occurring primarily on the plane parallel to the line of sight, such a rotational vortex structure would not be observed in the plane of the sky. ICM profiles in segmented regions show a tendency of blue-shift of 
% \color{red} $\sim$250~km~s$^{-1}$  \color{black} to the south and red-shift of  \color{red} 0--100~km~s$^{-1}$  \color{black} to the north. 
Significant blue-shifted motion at Reg~B (around the east edge of the cold front) and almost zero red shift in Reg-A (to the south) suggest an ICM flow following the BCG is rotating with some inclination to the line of sight, and the component in Reg-B is ``falling'' into the BCG potential from the far side.
The Chandra X-ray brightness and temperature maps show that Reg-B is likely a distinct structure in the sloshing core, with the lowest specific entropy \citep{2021MNRAS.504.2800I}. In simulations of sloshing (e.g., \cite{2006ApJ...650..102A}), the lowest-entropy gas is continuously separating itself from the rest of the moving core, falling towards the peak of the gravitational potential, overshooting it, and repeating the process, and our velocity findings appear consistent with this picture. 
\color{black}
%
%
%Chandra residual and temperature maps (e.g., \cite{2016MNRAS.461..684W,2021MNRAS.504.2800I}), %which 
% inferred that the sloshing motion also has a motion in the skyplane. Our XRISM/Resolve-based spectroscopic measurements, namely, the finding that the line of sight velocities relative to the BCG are close to zero across most of the core,\color{red} therefore provide the first direct confirmation in velocity space of the sky-plane geometry inferred by \citet{2021MNRAS.504.2800I}, \color{black} provide the first direct confirmation of this geometry in velocity space, supporting the picture inclined rotating low entropy ICM. % derived from X-ray morphology and thermodynamics.
Although the detailed geometry of the ICM flow is not clear, the ``spiral structure'' and the ``falling low entropy gas'' suggest that the sloshing, or ICM rotation in the central region, has some inclination to the skyplane.
\color{black}

%However, the observed bulk velocity difference between the center and the cold front in the Resolve data is less than XX~km~s${-1}$ within a 3σ confidence interval. This indicates that the sloshing motion of the low-entropy ICM in A2319 is occurring in a plane close to the plane of the sky.

%3.2章で要約したコールドフロント付近の視線上のBulk速度はBCG付近のBulk速度に近い。もしSloshingによって低エントロピーガスが角運動量を持って回転しているのであれば、最もエントロピーが低い中心のICMとコールドフロント近くのICMは3次元的に別の方向に動くはずである。シミュレーションではコールドフロントに伴うICMの運動は典型的にマッハ0.3-0.5と亜音速であることを示している。

%マイナー衝突の流体シミュレーション(Zuhone+2015)は、スロッシング運動の平面が視線上に沿った方向である場合、回転の中心とコールドフロント付近の低エントロピーICMは視線上に200-300km/s(マッハ0.2-0.3)の速度差があることを示している。A2319の8.5~keVの低エントロピーICMにとって、音速は、平均分子量$\mu=0.6$と比熱比$\gamma=5/3$を用いて、$C_{s}=({\gamma k_{B}T}/{\mu m_{p}})^{1/2}$=1500~km~s${-1}$である。もし、低エントロピーICMが視線上に回転していたとすると、視線上に300-400~km~s${-1}$の速度差が観測される。Resolveの中心とコールドフロントのBulk速度差は3σの信頼区間でXX~km~s${-1}$以下である。このことはA2319の低エントロピーICMのSloshing運動は、天球面上に近い平面で起こっていることを示している。

%ChandraのX線画像はSloshingによるBCGから反時計周りに約100~kpcの大きさを持つ渦構造の存在を示している。もし、Sloshingの運動が視線上にそった方向であれば回転による渦構造は天球面上で見られないはずであり、渦構造はSloshing運動が天球面上で起きていることを指示している。

%\color{blue}X線画像から反時計回りに回転していることを述べる。\color{black}

%\color{magenta}＊yes。

%\color{blue}BCGとBulk速度の関係。\color{black}

%\color{magenta}＊不要では？

%＊＊：全体的に良いと思います。英語化を進めしょう。\color{black}

\subsection{Development of turbulence near the cold front}
%コールドフロント付近の乱流の発達
The velocity dispersion map in figure~\ref{fig3:Resolve map} shows that the ICM near the 
southern part of the 
\color{black}
cold front (Reg-A) has a 
velocity dispersion
value of %350$\pm$80~km~s${-1}$,
$405^{+191}_{-96}$~km~s$^{-1}$,
\color{black}
%Actually, the values are higher to the south. 
%Slightly higher values are seen in the southern regions, 
which may reflect enhanced turbulence or projection effects in the downstream of the sloshing flow.
%which is clearly higher than that in other regions.
In observations of large-scale rotational motions such as sloshing, emission lines can be broadened due to the superposition of multiple ICM components with different line-of-sight velocities \citep{2003AstL...29..791I}. 
We cannot distinguish based on our data alone whether it is turbulence or shear; however, the Chandra image strongly suggests that the front at this location is oriented in the sky plane, with the velocity flows occurring in the sky plane, thus strong shear along the line of sight is less likely than turbulence. 
\color{black}
Of course, there can be two or more overlaid components with different line of sight velocities there, as discussed later. 
%As discussed in Section 4.1, it is considered that the low-entropy rotation occurs in a plane that is nearly aligned with the plane of the sky. Then, what might the significant velocity dispersion
%observed in the ``downstream'' of the sloshing flow indicate? 

%The simplest model is that all of this velocity dispersion is caused by isotropic turbulence. The turbulent Mach number of the ICM near the center is \color{red} $M_{3D}=\sqrt{3}\sigma_{v}/c_{s}\sim$0.2, which is the same value calculated for Abell 2029. \color{black} In contrast, the turbulent Mach number near the cold front is $M_{3D}$=0.4, indicating a relatively high subsonic speed. This suggests that turbulence may be developing over time within the sloshing flow. 
If we interpret the line broadening as isotropic turbulence, then its
\color{black}
turbulent Mach number
%near the center is $M_{3D} = \sqrt{3}\sigma_v / c_s \sim 0.2$, consistent with the value reported for Abell~2029. In contrast, the Mach number 
near the cold front reaches $M_{\rm {3D}} =0.49^{+0.23}_{-0.12}$, indicating relatively stronger subsonic turbulence. 
It corresponds to a nonthermal pressure (NT) fraction of 
\begin{equation}
\frac{P_{\rm NT}}{P_{\rm tot}} = \frac{M_{\rm 3D}^2}{M_{\rm 3D}^2 + 3/\gamma}=11.2^{+10.6}_{-4.7}\%, 
\end{equation}
assuming an adiabatic index of \(\gamma = 5/3\) (e.g., \cite{2019A&A...621A..40E}).
This may reflect continuous energy injection by sloshing-induced shear flows, or possibly an evolution from initially anisotropic motions toward a more isotropic turbulent state over time, as discussed in the context of the Coma cluster (e.g., \cite{2019NatAs...3..832Z,2012MNRAS.421.1123C}).
\color{black}

Such high velocity dispersion near the cold front may also be associated with eddy-like features produced by shear-induced instabilities. Previous studies based on Chandra images have suggested the presence of Kelvin Helmholtz instability (KHI) along the cold front boundary in A2319,
especially to the west of the segment of the front in Reg-A,
\color{black}
as indicated by image residual 
and temperature structures \citep{2021MNRAS.504.2800I}. If such KHI-induced eddies are present, they could introduce anisotropic velocity perturbations that are observed as enhanced dispersion when spatially averaged by the PSF of XMA. 
Such eddies would also drive turbulence.
\color{black}
This scenario is consistent with the observed broadening,
the variations of the line of sight velocity within Reg-A,
\color{black}
and the possible redistribution of kinetic energy into turbulence in the downstream region.
\color{black}

%Another possible interpretation is that, in the "downstream" of the sloshing flow, the ICM is diffusing perpendicularly to the flow direction, that is, along our line of sight. In either case, the kinetic energy inherent in the ICM would be converted into turbulence and lateral motion, which could lead to a decrease in the flow velocity in the downstream region. In fact, a slight decrease in the line-of-sight velocity is observed in the southeastern part of the sloshing region.
Alternatively, part of the observed velocity dispersion may result from bulk flow components projected along the line of sight. 
We divided Det-a into two sub-regions, east and west, but could not find statistically significant velocity difference.
Therefore, overall structure of the possible bulk motion is not clear. In the downstream of the sloshing flow, the ICM may be diffusing perpendicularly to the flow direction—that is, along our line of sight—leading to line broadening due to superposed velocity components. In either case, the kinetic energy of the flow would be redistributed into turbulence and lateral motions, potentially causing a decrease in the flow velocity downstream. Indeed, a slight drop in the line-of-sight velocity is observed in the southeastern region, supporting the presence of such redistribution.
Another possibility is the line of sight velocities diverging in the opposite directions at the nose of the front, as seen in simulations at certain stages and locations (e.g., \cite{2006ApJ...650..102A}), but this is unlikely at the current stage, with the spiral structure that we clearly see in the image, where most of the flow velocity inside the front should be in the sky plane along the front.
\color{black}
%図3の速度分散のマップはコールドフロント付近のICMの値は350$\pm$80~km~s${-1}$であり、それは他の領域の値と比較して明らかに大きいことを示している。Sloshingのような大きなスケールでの回転運動の観測では、視線速度の異なる複数のICM成分の重ね合わせによって輝線が広がることがある(Churazov+2003)。XRISM/Resoveは視野に対する計器点広がり関数(PSF)は50\%を超えていて、空間的なスペクトルの混合が起こりやすい。

%4.1章で議論したようにこの低エントロピーの回転はほぼ天球面上に近い面で起こっている。そのため、回転運動による視線上に異方的な速度差を持つICMの混合はほとんどなく、これらの速度分散がすべて等方性乱流に起因するものであるいと考えられる。この場合、中心付近のICMの乱流マッハ数は$M_{3D}=\sqrt{3}\sigma_{v}/c_{s}$=0.2であり、Abell2029で計算された値と同じ値である。一方でコールドフロント付近のICMの乱流マッハ数は$M_{3D}$=0.4と非常に高い亜音速であることを意味している。\\
%\color{magenta}→改訂案＊＊：4.1章で議論したようにこの低エントロピーの回転はほぼ天球面上に近い面で起こっている。sloshingのflowの「後方」で乱流が大きく見えることは何を意味するのだろうか？もっとも単純なモデルは、これらの速度分散がすべて等方性乱流に起因するものであると仮定することである。中心付近のICMの乱流マッハ数は$M_{3D}=\sqrt{3}\sigma_{v}/c_{s}$=0.2であり、Abell2029で計算された値と同じ値である。一方でコールドフロント付近のICMの乱流マッハ数は$M_{3D}$=0.4と非常に高い亜音速であることを意味している。このことから、sloshingのflowの中で、乱流が時間と共に発達している可能性がある。もう一つの考え方は、sloshing のflowの後方で、flow方向に垂直に、すなわち我々の視線方向にICMが拡散している可能性である。いずれの場合も、ICMに内在する運動エネルギーが乱流や横の動くに変換されることから、例えばsloshing のflow速度がflow後方で低下する可能性がある。実際、sloshingの南東部ではわずかにline of sight velocityが低下している傾向があり、注目される。\color{black}

%\color{blue}乱流成長の一般論。\color{black}
%Zuhoneのシミュレーションはコールドフロントの内側に位置するスロッシングの冷たいガス成分は、輝線が著しくブロード化する領域であることを示している。
%＊＊：これは何をエネルギー・運動量の基として実施されているのだろうか？

\subsection{High redshift clump (A2319B)}
%高赤方偏移のクランプ
As stated in the introduction, A2319 hosts an isolated optical galaxy group (A2319B) located approximately 10 arcminutes northwest of the center, with an average velocity of 18636~km~s$^{-1}$ or $z = 0.06212$ \color{black} \citep{1995AJ....110...32O},
redshifted by 3000~km~s$^{-1}$ from the BCG.
Identifying ICM emission from A2319B is important to understand the merger geometry around this cluster.
\citet{1996ApJ...465L...1M} used ASCA data 
%to discover 
and found that \color{black} the region located at A2319B is relatively cooler compared to the surrounding areas. 
%\citet{2009PASJ...61.1293S} found, through Suzaku/XIS analysis, that this cold region has a high metal abundance, but no evidence of ICM motion in the subgroups with a velocity difference of 3000~km~s$^{-1}$ was observed. 
Using Suzaku XIS data, \citet{2009PASJ...61.1293S} further discovered that this cold region has a higher elemental abundance compared to the surrounding regions. On the other hand, redshift measurements showed no significant signs of redshifted ICM compared to the surrounding areas.
This non-detection is likely attributable to the limited energy resolution of the CCD detectors, which makes it challenging to resolve redshift differences of a few thousand km~s$^{-1}$ amidst line blending and calibration uncertainties.
\color{black}

Although the Resolve field of view does not cover that region,
the galaxies from that subcluster do spread across the whole cluster, and 
\color{black}
we searched for an ICM component redshifted by 3000~km~s$^{-1}$ using data from Resolve, which offers more than 30 times higher energy resolution than XIS. We added a model component with a velocity of 18636~km~s$^{-1}$ to the baseline model listed in table~\ref{tab1:Fit parameter} and performed spectral fitting. The velocity dispersions were fixed to the values given in table~\ref{tab1:Fit parameter}. 
In the lower panel of figure~\ref{fig2:FoV spec}, the expected line centers of He-like Fe K$\alpha$ complex (w, x, y, z) and H-like Fe Ly$\alpha_{1,2}$ are indicated by blue lines.
\color{black}
In the fitting results, the upper limit of the flux corresponding to the component at 18636~km~s$^{-1}$ was calculated to be 2.66$\times$10$^{-16}$~erg~s$^{-1}$~cm$^{-2}$~sr$^{-1}$ (0.5--10.0~keV)
\color{black}
at a 90\% confidence interval. %This value falls within the range of systematic errors due to statistical fluctuations, and therefore, it can be concluded that the component at 18636~km~s$^{-1}$ was not detected in this observation.
This value can be considered as statistical fluctuations, and it can be concluded that the component at 18636~km~s$^{-1}$ was not detected in this observation.

\subsection{High temperature ICM in northern region}

The temperature map in figure~\ref{fig3:Resolve map} shows that Reg-E has an average temperature of approximately 11~keV, which is higher than that of the surrounding regions. 
Chandra data \citep{2021MNRAS.504.2800I} also show that Reg-E includes another cold front, distinct from the main one on the eastern side of the core. This secondary front is located at a much smaller off-center distance and exhibits a significant temperature gradient. It is a site where hot gas from larger radii is driven inward by sloshing motions, approaching the cluster center and eventually mixing with the cooling core gas. This process provides a heat influx into the core and may counteract runaway cooling 
\citep{2004ApJ...612L...9F,2010ApJ...717..908Z}.
\color{black}
%Previous temperature maps from Chandra and XMM-Newton revealed that the cold front region, characterized by a temperature around 8~keV, is enveloped by hotter ICM, particularly in the northwest, where a belt-like structure with temperatures of 11–13~keV is observed near the BCG \citep{2004ApJ...605..695G,2021MNRAS.504.2800I}. This high-temperature region is partly overlapping with Reg~E and is associated with higher entropy than the surrounding gas \citep{2021MNRAS.504.2800I}.
%Numerical simulations of gas sloshing suggest that as low-entropy ICM rotates inward, high-entropy, high-temperature gas from outer regions can flow in and mix with the colder ICM, leading to entropy flattening \citep{2010ApJ...717..908Z}. The high-temperature region in the northwest of A2319 may be undergoing such mixing with the low-temperature core.

%ここを変える・領域Eがそもそもredshiftしている。2成分が混ざっているとすると...
%The bulk velocity map indicates that the high-temperature region in the northwest is redshifted at an overall velocity of 100~km~s$^{-1}$ relative to the BCG. The temperature is approximately 9.8~keV on average, which is lower compared to the 11~keV suggested by the Chandra temperature map. This suggests that the low- and high-temperature components may be mixing in the spectrum. % due to the PSF of Resolve. 

\begin{figure}[htb]
 \begin{center}
  \includegraphics[width=8cm]{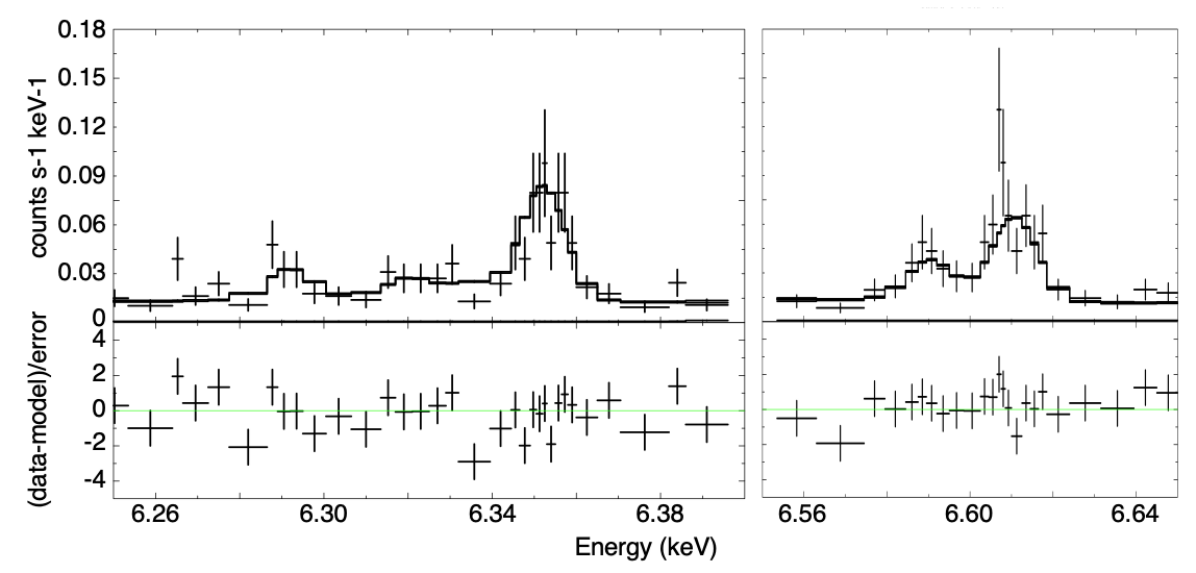} 
 \end{center}
\caption{Zoom-in of the Resolve spectrum around the He-like (left) and H-like (right) Fe lines of the Det-e'. Data are binned by a factor of 2~eV for display.
{Alt Text: Two line graphs. In the left panel, the x axis shows the energy from 6.25 to 6.40 kilo electron volt. In the lower right panel, the x axis shows the energy from 6.55 to 6.65 kilo electron volt. The y axis shows the count from 0.0 to 0.18 counts per second and per kilo electron volt, and the residuals of minus 5 to 5 in lower part.}}
\label{fig4:Hot region}
\end{figure}

It is interesting to try to detect any velocity differences within this region, given that it includes gas inside and outside the cold front, which has different temperatures (with the cooler, denser gas located inside the cold front).
\color{black}
Reg-E is redshifted by about 90~km~s$^{-1}$ relative to the BCG. The spectrum of Det-e' (see figure~\ref{fig4:Hot region}) shows that 
\color{black}
the position of the He-like Fe lines (6.2–6.4 keV),
which come from gas in a broad range of temperatures,
\color{black}
matches well 
between the model and the data,
%with the spectrum, 
while
%but the position of 
that of the H-like Fe lines (6.5–6.7 keV), 
which originate from hotter gas phases, is more redshifted in the data.
\color{black}
When fitting the spectrum with the temperature fixed,
the bulk velocity relative to the BCG
is found to be 
$37.7_{-38.6}^{+39.7}$~km~s$^{-1}$(6.2--6.4~keV) 
and 
$92.6_{-33.5}^{+40.1}$~km~s$^{-1}$(6.5--6.7~keV),
respectively.
The difference
suggests that the low-temperature and high-temperature ICM components have different velocities and these components are mixed in the spectrum.
%These results suggest the presence of a velocity difference between the cooler ICM in the cluster core and the high-entropy, hotter ICM mixing in from the outer regions. 
%However, the current dataset lacks sufficient statistical significance to clearly separate these two spectral components. 
%Therefore, to enable a more detailed physical interpretation, deeper observations of the same region or additional pointings, particularly toward the northwest, where the hotter component is expected to be more prominent, will be required in future studies.
The current dataset, however, lacks sufficient statistical significance to clearly separate these two spectral components. Additional XRISM pointing(s) around the high-entropy ICM region will solve this in the future.

\section{Conclusions}
We presented results from the XRISM observations of 
%Abell 2319, 
the core of A2319, 
one of the X-ray brightest nearby 
hot
clusters, focusing on its kinematic structure.
The ICM temperature was found to be approximately 8~keV across the cluster core, with a prominent cold front structure and a region of elevated temperature ($\sim$10 keV) in the northwest.
%Spatially resolved spectroscopy revealed that the southern region relative to the BCG exhibits a blue shift of about 100~km~s$^{-1}$, although no significant velocity gradient was detected within the region. 
The bulk velocity  around the BCG is consistent with that of the BCG within 40~km~s$^{-1}$.
A blueshift of up to $\sim$230~km~s$^{-1}$ is observed along the eastern edge of the cold front, located southeast of the BCG,
where the gas with the lowest specific entropy is found,
\color{black}
while the region farther south along the cold front shows only a modest redshift. However, in the southern region inside that front, the velocity dispersion increases to $\sim$400~km~s$^{-1}$, suggesting the development of turbulence. These features indicate that the ICM is undergoing sloshing motion with a certain inclination angle relative to the plane of the sky, following the motion of the BCG.
We also investigated the possible presence of high-redshift ICM associated with the subcluster A2319B but found no significant detection within the current field of view. 

Overall, these results provide new insights into the internal dynamics of A2319, highlighting sloshing motions that occur with a moderate inclination to the plane of the sky. 
The gas phases with different specific entropies move independently of each other, as expected from simulations of sloshing. 
\color{black}
They also contribute to our understanding of the cluster's three-dimensional geometry and the role of turbulence in shaping the thermal and kinematic structure of the ICM.
Deeper observations and additional pointings will be essential to further explore the detailed nature of the high-temperature gas and the turbulent properties of the sloshing flow.
\color{black}

\begin{ack}
We gratefully acknowledge the hard work over many years of all of the engineers and scientists who made the XRISM mission possible. Part of this work was performed under the auspices of the U.S. 
This work was supported by the JSPS Core-to-Core Program, JPJSCCA20220002. The material is based on work supported by the Strategic Research Center of Saitama University.
Department of Energy by Lawrence Livermore National Laboratory under Contract DE-AC52-07NA27344. The material is based upon work supported by NASA under award numbers 80GSFC21M0002 and 80GSFC24M0006. This work was supported by JSPS KAKENHI grant numbers 
JP19K14762, JP19K21884, JP20H00157, JP20H01946, JP20H01947, JP20H05857, JP20K04009, JP20K14491, JP20KK0071, JP21H01095, JP21H04493, JP21K03615, JP21K13958, JP21K13963, JP22H00158, JP22H01268, JP22K03624, JP23H00121, JP23H00151, JP23H01211, JP23H04899, JP23K03454, JP23K03459, JP23K13154, JP23K20239, JP23K20850, JP23K22548, JP24H00253, JP24K00638, JP24K00672, JP24K00677, JP24K17093, JP24K17104, and  JP24K17105.
This work was supported by NASA grant numbers 80NSSC22K1922, 80NSSC18K0978, 80NSSC18K0988, 80NSSC18K1684, 80NSSC20K0733, 80NSSC20K0737, 80NSSC20K0883, 80NSSC23K0650, 80NSSC23K1656, 80NSSC24K0678, 80NSSC24K1148, 80NSSC24K1774.
LC acknowledges support from NSF award 2205918. CD acknowledges support from STFC through grant ST/T000244/1. LG acknowledges financial support from Canadian Space Agency grant 18XARMSTMA. NO acknowledges partial support by the Organization for the Promotion of Gender Equality at Nara Women’s University. MS acknowledges the support by the RIKEN Pioneering Project Evolution of Matter in the Universe (r-EMU) and Rikkyo University Special Fund for Research (Rikkyo SFR). AT and the present research are in part supported by the Kagoshima University postdoctoral research program (KU-DREAM). SY acknowledges support by the RIKEN SPDR Program. TY acknowledges support by NASA under award number 80GSFC24M0006. IZ acknowledges partial support from the Alfred P. Sloan Foundation through the Sloan Research Fellowship. SE acknowledges the financial contribution from the contracts Prin-MUR 2022 supported by Next Generation EU (M4.C2.1.1, n.20227RNLY3 The concordance cosmological model: stress-tests with galaxy clusters), ASI-INAF Athena 2019-27-HH.0, “Attività di Studio per la comunità scientifica di Astrofisica delle Alte Energie e Fisica Astroparticellare” (Accordo Attuativo ASI-INAF n. 2017-14-H.0), and from the European Union’s Horizon 2020 Programme under the AHEAD2020 project (grant agreement n. 871158). LL acknowledges the financial contribution from the INAF grant 1.05.12.04.01.
YO would like to take this opportunity to thank the “Nagoya University Interdisciplinary Frontier Fellowship” supported by Nagoya University and JST, the establishment of university fellowships towards the creation of science technology innovation, Grant Number JPMJFS2120.
Y.O. was supported by the Sasakawa Scientific Research Grant from The Japan Science Society.
\end{ack}

\appendix
\section{Energy scale calibration and uncertainty}
\label{appendixA}
The gain of the XRISM Resolve microcalorimeter detectors is affected by its environment including the temperature of the 50~mK heat sink, bolometric effects of the detectors themselves due to thermal radiation from the instrument thermal shields, and the temperature of the amplifier and bias electronics in the warm instrument electronics. The net effect is a detector gain, and thus energy scale error, that is time dependent. XRISM carries multiple on-board calibration sources to monitor the time dependent gain and to reconstruct the time dependent energy scale using a non-linear correction method described in \citet{2016JLTP..184..498P}. 

Current XRISM observations obtain gain fiducials by rotating $^{55}$Fe x-ray sources on the Resolve filter wheel into the instrument field of view for 30 minute intervals strategically placed during each observation. The gain correction scheme, described in 
\color{black} \cite{Porter2025} and \cite{Sawada2025}, \color{black} is highly effective in reconstructing the energy scale, typically to within 0.1--0.2~eV at 6~keV. The gain fiducial intervals are closely spaced around events that disturb the gain which include recycling of the 50~mK refrigerator and large scale temperature changes of the instrument electronics due to changes in the spacecraft orientation after slewing to a new target. Outside of these large disturbances, the time dependent instrument gain is extremely linear and is only sparsely sampled with gain fiducials approximately every 20 hours in the standard XRISM/Resolve observing strategy. Finally, the Resolve instrument includes a calibration pixel that is part of the microcalorimeter array but located just outside the field of view. The calibration pixel is continuously illuminated by a heavily collimated $^{55}$Fe x-ray source. In standard XRISM/Resolve observations, the calibration pixel is used to measure the efficacy of the energy scale reconstruction by comparing the Mn Ka fiducial line offset and resolution during the observation, but only gain corrected using the sparsely sampled fiducial intervals used for the main observation. These metrics are a standard product of the XRISM pipeline.

\begin{figure*}[htpb]
 \begin{center}
  \includegraphics[width=17.0cm]{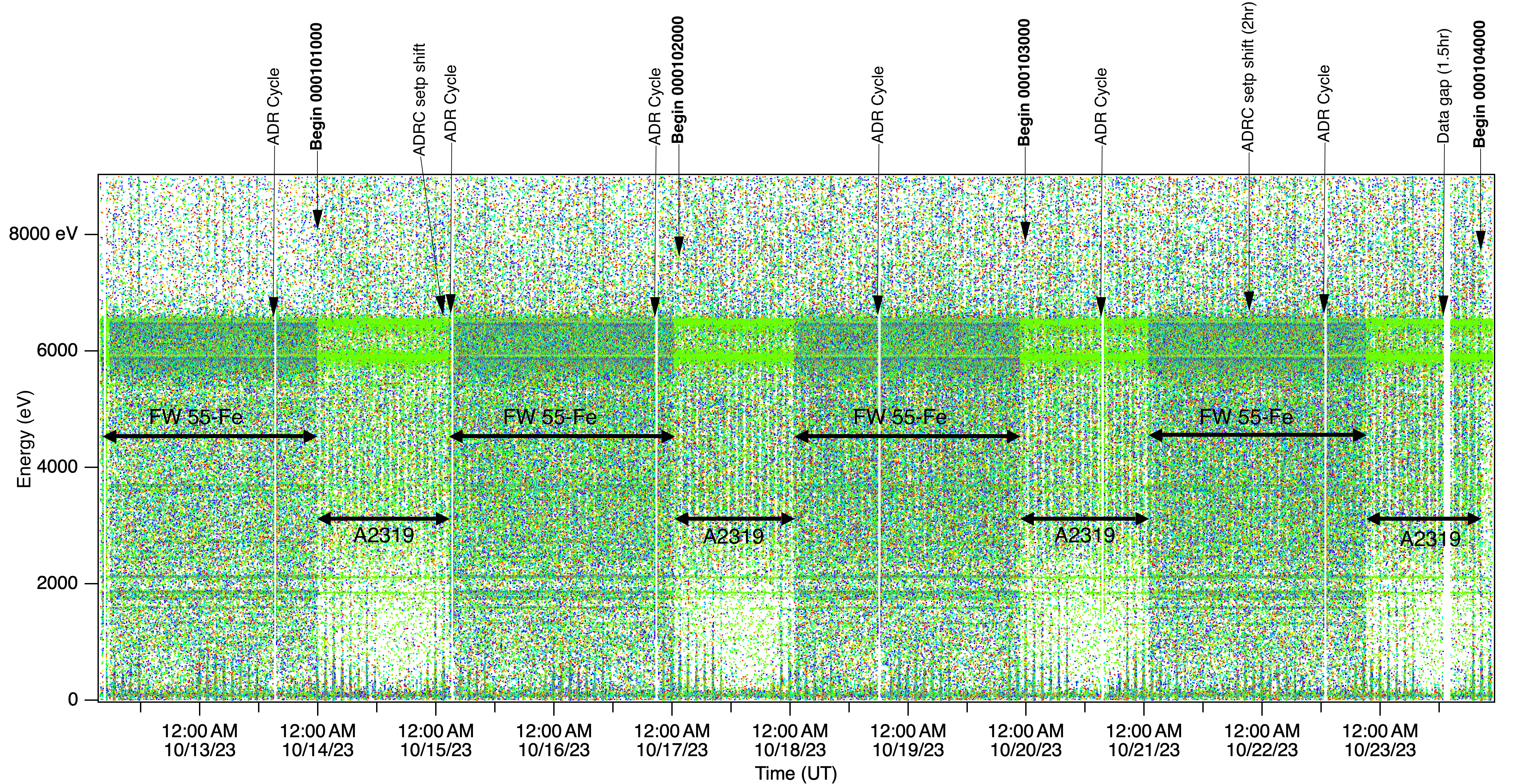} 
 \end{center}
\caption{An overview of the A2319 observations where each ``dot" represents an x-ray event color coded by pixel. The start of each OBSID is indicated along with the positions of the ADR refrigerator recycles. Further, the periods when the filter wheel $^{55}$Fe x-ray sources were in the field of view are indicated along with the unobscured observations of Abel 2319. The light green points at 5.9 and 6.4 keV are the continuously illuminated calibration pixel.
{Alt text: Two-dimensional scatter plot of Resolve photon events. In the panel, x-axis is time, from 12 to 24 October 2023, and y-axis is photon energy after gain calibration, from 0~eV to 9000~eV. There are four periods with $^{56}$Fe irradiation and another four without. The latter can be used for our analyses. }}
\label{fig2:Obs_overview}
\end{figure*}

As the
%The 
observations of A2319 %presented here were %observed 
were carried out very early in the XRISM commissioning phase, before the modern observing strategy was implemented, %Thus 
\color{black}
the standard, sparsely sampled gain fiducial methodology was not employed. However, during this time, the Resolve instrument team was benchmarking the on-orbit instrument performance using the same $^{55}$Fe filter wheel sources now used for gain reconstruction. An event history plot for the A2319 ``core" observations is shown in figure~\ref{fig2:Obs_overview}. During the A2319 observations, there are long periods with and without the filter wheel $^{55}$Fe sources in the field of view. While the periods with the $^{55}$Fe sources are not usable for science, they provide adequate gain calibration fiducials for the periods without the sources in the field of view. This required some careful handling of the observation data because there were subtleties that are not properly handled by the standard Resolve pipeline gain calibration methodology. For example, the beginning fiducial for OBSID 000101000 actually occurs during the previous OBSID, similarly for OBSIDs 000102000 and 000103000. There is also a small 50~mK set point error in the 50~mK temperature controller during OBSID 000101000 that was corrected by cross-calibration with the cal pixel (the set point error occasionally occurred during passages through the South Atlantic Anomaly and was permanently fixed later in commissioning). During OBSID 000103000 there was a recycle event of the Adiabatic Demagnetization Refrigerator (ADR) that cools the 50~mK detector heat sink. This heavily disturbs the detector gain. In this case, the detector gain was forward extrapolated from the earlier fiducial interval up to the ADR recycle since this is firmly within the linear gain drift period for the detectors. The period after the ADR recycle was back-extrapolated from the following gain fiducial, but only during the linear portion of the gain history which begins 6 hours after the ADR recycle \citep{Porter2025}. Finally, the last portion of OBSID 000103000 has no ending fiducial measurement, but is firmly within the linear gain drift period, and was thus forward extrapolated from the prior fiducial region.

In all cases, the calibration pixel was analyzed using the same time-dependent gain correction method described above for the main array. However, the calibration pixel is continuously illuminated, so we fit the Mn Ka line from the $^{55}$Fe source to track the energy resolution and the line shift as a measure of the efficacy of the energy scale reconstruction. This is the same methodology used in all Resolve observations. This gave cal pixel resolutions of 4.45 $\pm$ 0.02, 4.45 $\pm$ 0.02, and 4.46 $\pm$ 0.02~eV FWHM at 5.9~keV for OBSIDs 000102000, 000102000, and 000103000 respectively, which is exactly inline with the ensemble of XRISM/Resolve observations \citep{Porter2025,Kelley2025,Leutenegger2025}. Similarly, the energy scale errors were $-0.41 \pm$ 0.01, $0.14 \pm$ 0.01, and $0.04 \pm$ 0.01~eV for the same OBSIDs. Using the standard resolve method for estimating the energy scale systematic uncertainties \citep{Eckart2025}, we add the cal pixel uncertainties in quadrature with estimates of the underlying energy scale uncertainty of 0.3 eV in the band 5.4--9~keV, and 1~eV outside of that band. This yields energy scale uncertainties for these observations of 0.51, 0.33, and 0.30~eV in the energy band of 5.4--9~keV, respectively. 

%\section{NoSSM}

Because these A2319 observations need special care on gain handling, we present the ICM parameter maps before applying the SSM correction in 
%figure~\ref{fig6:Resolve map noSSM}. 
table~\ref{tab2:Fit parameter in detector regions}. 
\color{black}
As expected, tendencies of the parameters are very similar to those of figure~\ref{fig3:Resolve map}. SSM corrects the smearing effect caused by the modest PSF of XRISM, and therefore differences between the parameters are enhanced after correction.

\color{black}

\begin{table*}[h!]
\begin{center}
\tbl{Best-fit Parameters of the detector regions in 3.0--9.5~keV energy range. }{%
\resizebox{0.95\textwidth}{!}{%
\begin{tabular}{@{}lcccccccc@{}}
\hline\noalign{\vskip3pt}
\multicolumn{1}{l}{Parameters} & Det-a$^{*}$ & Det-b & Det-d & Det-b' & Det-c' & Det-d' & Det-e'  \\ [2pt]
\hline\noalign{\vskip3pt}
$kT$~(keV) & $7.66^{+0.35}_{-0.37}$ & $7.72^{+0.49}_{-0.49}$ & $8.66^{+0.69}_{-0.60}$ & $8.42^{+0.42}_{-0.35}$ & $8.50^{+0.32}_{-0.34}$ & $9.26^{+0.45}_{-0.40}$ & $9.84^{+0.49}_{-0.48}$  \\
Abundance~($Z_{\rm{\odot}}$) & $0.35^{+0.04}_{-0.04}$ & $0.39^{+0.05}_{-0.05}$ & $0.34^{+0.05}_{-0.05}$ & $0.39^{+0.04}_{-0.03}$ & $0.37^{+0.03}_{-0.03}$ & $0.36^{+0.04}_{-0.04}$ & $0.40^{+0.04}_{-0.04}$  \\
Redshift ($\times$10$^{-2}$) & $5.4435^{+0.01852}_{-0.01821}$ & $5.4367^{+0.01351}_{-0.01321}$ & $5.4667^{+0.01302}_{-0.0106}$ & $5.4155^{+0.01116}_{-0.0102}$ & $5.4689^{+0.0084}_{-0.0093}$ & $5.4468^{+0.0105}_{-0.0112}$ & $5.4817^{+0.01000}_{-0.0082}$ \\
Relative Velocity (km~s$^{-1}$) & $-60.0^{+55.6}_{-54.6}$ & $-76.3^{+40.5}_{-39.6}$ & $+13.8^{+39.0}_{-31.6}$ & $-140.0^{+33.5}_{-30.6}$ & $+20.1^{+25.2}_{-28.0}$ & $-46.0^{+31.4}_{-33.7}$ & $+58.6^{+30.0}_{-24.5}$ \\
Velocity dispersion (km~s$^{-1}$) & $312^{+54}_{-44}$ & $232^{+32}_{-31}$ & $147^{+37}_{-40}$ & $246^{+36}_{-32}$ & $206^{+24}_{-23}$ & $195^{+41}_{-38}$ & $193^{+28}_{-26}$ \\
C-stat/$d.o.f$ & $7652/12994$ & $9748/12994$ & $9103/12994$ & $20243/25992$ & $20675/25992$ & $19058/25992$ & $20627/25992$ \\
[2pt]
\hline\noalign{\vskip3pt}
\end{tabular}
}} % resizebox ここまで
\label{tab2:Fit parameter in detector regions}
\begin{tabnote}
{\hbox to 0pt{\parbox{170mm}%78mm}
{\footnotesize
\par\noindent
$*$ Spectral fitting results obtained without applying SSM. Det-a, b and d are in the data of OBSID:000101000, almost matching with Reg-A, B, and D.
Det-b', c', d' and e' are in the data of OBSID:000102000 and 000103000, almost matching Reg-B, C, D and E. \color{black}
}\hss}} 
\end{tabnote}
\end{center}
\end{table*}

%%%
% See the manual for the detail.
%%%

\bibliographystyle{pasj} % style pasj.bst
\bibliography{pasj} % your references 

\end{document}